\documentclass{jfmc}
\usepackage{graphicx}
\usepackage{epstopdf, epsfig}

\usepackage{amsmath}

\shorttitle{Closures in the volume-filtered framework}
\shortauthor{M.~Hausmann, V.~Ch\'eron, F.~Evrard and B.~van~Wachem}

\title{Study and derivation of closures in the volume-filtered framework for particle-laden flows}

\author{Max~Hausmann\aff{1},
  Victor~Ch\'eron\aff{1}, Fabien~Evrard\aff{2}
 \and Berend~van~Wachem\aff{1}}

\affiliation{\aff{1}Chair of Mechanical Process Engineering, Otto-von-Guericke-Universität Magdeburg, Universitätsplatz 2, 39106 Magdeburg, Germany\aff{2}Department of Aerospace Engineering, University of Illinois Urbana-Champaign, 104 South Wright Street, Urbana, IL 61801, United States}

\begin{document}

\maketitle

\begin{abstract}
The volume-filtering of the governing equations of a particle-laden flow allows for simulating the fluid phase as a continuum and accounting for the momentum exchange between the fluid and the particles by adding a source term in the fluid momentum equations. The volume-filtering of the Navier-Stokes equations allows to consider the effect that particles have on the fluid without further assumptions, but closures arise of which the implications are not fully understood. It is common to either neglect these closures or model them using assumptions of which the implications are unclear. In the present paper, we carefully study every closure in the volume-filtered fluid momentum equation and investigate their impact on the momentum and energy transfer dependent on the filtering characteristics. We provide an analytical expression for the viscous closure that arises because filter and spatial derivative in the viscous term do not commute. An analytical expression for the regularization of the particle momentum source of a single sphere in the Stokes regime is derived. Furthermore, we propose a model for the subfilter stress tensor, which originates from filtering the advective term. The model for the subfilter stress tensor is shown to agree well with the subfilter stress tensor for small filter widths relative to the size of the particle. We show that, contrary to common practice, the subfilter stress tensor requires modeling and should not be neglected. For small filter widths, we find that the commonly applied Gaussian regularization of the particle momentum source is a poor approximation of the spatial distribution of the particle momentum source, but for larger filter widths the spatial distribution approaches a Gaussian. Furthermore, we propose a modified advective term in the volume-filtered momentum equation that consistently circumvents the common stability issues observed at locally small fluid volume fractions and identify inconsistencies in previous studies of the phase-averaged kinetic energy of the volume-filtered fluid velocity. Finally, we propose a generally applicable form of the volume-filtered momentum equation and its closures based on clear and well founded assumptions and propose guidelines for point-particle simulations based on the new findings.
\end{abstract}


\section{Introduction}
Capturing the detailed flow physics of natural and industrial scale phenomena with numerical simulations has always been, and still is, out of reach due to the tremendous amount of computational resources required \citep{Tenneti2014}. Approaches to capture only the relevant features of the involved physics are frequently applied as a useful remedy. The most important contributions are the decomposition of the flow quantities into a temporal mean and a fluctuation, proposed by \citet{Reynolds1895}, that forms the basis of the Reynolds averaged Navier-Stokes (RANS) models, and the decomposition into a large scale and small scale contribution, which is exploited in large eddy simulations (LES) \citep{Smagorinsky1963,Lilly1966,Schumann1975}. The large scale quantities are typically associated to filtered field quantities and the small scales are the remaining subfilter scales \citep{Sagaut2005}. In numerical simulations, the fluctuating or subfilter contributions often require the majority of the computational resources, as they need a high spatial and temporal resolution, although their (energetic) contributions to the physical configuration are often minor. The unknown fluctuating or subfilter scales have an effect on the mean or filtered quantities. For single-phase flows, this effect is represented by an additional stress term, which has to be modeled. \\
An additional difficulty compared to single-phase flows arises in particle-laden flows, as the fluid quantities are not continuous, i.e., they are not defined inside the particle. Consequently, the temporal averaging or spatial filtering of single-phase flows required adaption \citep{Drew1999}. To that end, a particularly interesting approach was derived by \citet{Anderson1967}, referred to as the volume-filtering approach. The volume-filtering approach is closely related to the filtering of a single-phase flow, with the exception of excluding the volumes occupied by the particles in the convolution integral. Therefore, the volume-filtering yields an instantaneous and continuous representation of fluid and fluid flow quantities and converges towards the LES filtering far away from particles. Applying the volume-filtering to the governing fluid equations, the Navier-Stokes equations (NSE), results in a set of equations that is not closed, as a number of terms that contain unknown subfilter quantities arise. There are two reasons for these unclosed terms to appear: (\romannumeral 1) Volume-filtering the non-linear advective term does not result in the equivalent advective term for the filtered velocity, but entails an additional term, i.e.,  the subfilter stress tensor, which also arises in the case of single-phase flow filtering. (\romannumeral 2) Volume-filtering and spatial derivatives are not commutable, which causes two additional unclosed terms, a stress closure that can be interpreted as the spatial distribution of the particle momentum source, and a viscous closure originating from the second spatial derivative in the viscous term. \citet{Capecelatro2013} presented a volume-filtered framework for flows with a large particle volume fraction where the subfilter stress tensor is omitted without much corroboration. \\
In the literature, the viscous closure is sometimes modeled as an additional viscosity, which is justified by the observation that particles increase the effective or apparent viscosity of the flow \citep{Zhang1997a,Patankar2001a,Gibilaro2007,Capecelatro2013,Arolla2015}, or completely neglected \citep{Balachandar2019,Subramaniam2022}. However, a direct relation between the effective viscosity and viscous closure has not been shown, nor does a theoretical or empirical basis exists to generally neglect the viscous closure.  \\
The modeling of the second closure, i.e., the subfilter stress tensor, is particularly difficult because it contains the contributions of both the unresolved turbulence and the non-linear effect of the particles (sometimes referred to as "pseudo-turbulence" \citep{Mehrabadi2015}). In the context of filtering, only few attempts have been made to model the subfilter stress tensor. \citet{Shallcross2020} proposed a transport equation for the pseudo-turbulent kinetic energy to reconstruct the subfilter stress tensor for compressible particle-laden flows. Under the assumption that a point-particle DNS, that fully resolves the turbulence but treats the particles as points, captures the particle-turbulence interactions sufficiently well, an algebraic eddy viscosity model \citep{Yuu2001} and a one-equation model based on a transport equation for the subfilter kinetic energy \citep{Hausmann2023} have been proposed for LES. \\
The problem of modeling the third closure, i.e., the closure related to the particle momentum source, is twofold: (\romannumeral 1) The volume integral of the particle momentum source closure equals the negative hydrodynamic force that the particle experiences, and is typically modeled with empirical correlations. There are several intensely studied aspects of this force, such as the division into different contributions (drag, lift, added mass, history) \citep{Maxey1983}, non-sphericity of the particles \citep{Holzer2008,Zastawny2012c,Cheron2023b}, the prediction of the undisturbed fluid velocity at the particle center to enable the use of empricial force correlations \citep{Horwitz2016,Ireland2017,Horwitz2018,Balachandar2022,Evrard2020a}, and the effect of disturbances of neighboring particles on the hydrodynamic force \citep{Akiki2017,Balachandar2020,Lattanzi2022,vanWachem2023a}. (\romannumeral 2) Much less studied than the magnitude of the hydrodynamic forces, however, are the distribution or regularization of the hydrodynamic forces. The shape of the regularization of the hydrodynamic forces typically depends on the ratio between the particle size and the filter width. In numerical simulations, where the equations governing the flow are discretized on a numerical grid, the filter width is related to the size of a numerical grid cell. For small particles and large numerical grid cells, the particle momentum source is applied to the numerical grid cell that the center of the particle occupies, which essentially corresponds to a top-hat function and is known as the particle source in cell (PSIC) method \citep{Crowe1977}. As the particle size approaches the size of the numerical grid cell, the particle momentum source is regularized with a Gaussian or a polynomial approximation of a Gaussian \citep{Maxey2017,Maxey2001,Evrard2020a,Poustis2019,Keane2023}, which can be theoretically justified in the volume-filtered framework under the assumption that the filter width is much larger than the particle. It is unknown, however, how large the filter width has to be and how the shape of the regularization changes if the filter width is smaller. \\
In the present paper, we address all of the three closure problems mentioned above. We derive a novel analytical expression for the viscous closure that is valid without further restrictions and necessary to ensure Galilean invariance of the volume-filtered momentum equation. The influence of the subfilter stress tensor on the energy and momentum balance for different filter widths is studied and we propose a model that is derived from differential filtering. Furthermore, we derive an analytical expression for the regularization of the particle momentum source in the Stokes limit and compare it to the commonly used Gaussian for different filter width to particle size ratios. \\
A common simplification for dilute regimes, i.e., large fluid volume fractions or small particle volume fractions, is to assume a constant fluid volume fraction, such that the governing fluid equations are equal to the single-phase NSE with an additional particle momentum source term \citep{Maxey2017}. With this assumption, an equation for the phase-averaged fluid kinetic energy can be derived \citep{Xu2007a,Subramaniam2014,Mehrabadi2015}, which we show to be an oversimplification for large particle volume fractions or small fluid volume fractions. We extend this equation, such that it is valid regardless of the volume fraction. \\
The remainder of this paper is organized as follows: In section \ref{sec:VFEL}, the volume-filtering of the Navier-Stokes equations is introduced and the closures are described. Furthermore, we propose a new form of the advective term and discuss the aspect of Galilean invariance. In section \ref{sec:closures}, we discuss each closure term individually and provide analytical expressions or propose models for them. We study the impact of the closures on the energy transfer between the particles and the fluid and discuss the necessity of modeling them in section \ref{sec:evaluationclosures} by means of explicitly volume-filtered results from particle-resolved simulations. A consistent expression for the phase-averaged fluid kinetic energy equation is derived in section \ref{sec:fluidKE} and guidelines for the choice of the filter width and the implementation of the proposed closures are given in section \ref{sec:guidelines} before the paper is concluded in section \ref{sec:conclusions}.



\section{Volume-filtered Euler-Lagrange equations}
\label{sec:VFEL}
\subsection{Governing equations}
We consider the motion of a single or multiple rigid and non-rotating particles immersed in an incompressible Newtonian fluid flow with constant dynamic viscosity $\mu_{\mathrm{f}}$ and density $\rho_{\mathrm{f}}$. Outside of the particles, the fluid motion is governed by the NSE
\begin{align}
    \dfrac{\partial u_i}{\partial x_i} &= 0, \\
   \rho_{\mathrm{f}}\dfrac{\partial u_i}{\partial t} + \rho_{\mathrm{f}} \dfrac{\partial u_i u_j}{\partial x_j} &= - \dfrac{\partial p}{\partial x_i} + \mu_{\mathrm{f}} \dfrac{\partial}{\partial x_j}\left(\dfrac{\partial u_i}{\partial x_j}+\dfrac{\partial u_j}{\partial x_i}\right),
\end{align}
where $u_i$ is the fluid velocity vector and $p$ the pressure. At the boundary of the particle with the index $q$, $\partial \Omega_{\mathrm{p},q}$, the no-slip boundary condition is assumed:
\begin{align}
\label{eq:noslip}
    u_i|_{\partial \Omega_{\mathrm{p},q}} = v_{q,i},
\end{align}
where $v_{q,i}$ is the velocity of the particle with the index $q$. The particle motion is governed by Newton's second law:
\begin{align}
    \rho_{\mathrm{p},q} V_{\mathrm{p},q} \dfrac{\mathrm{d}v_{q,i}}{\mathrm{d} t}=F_{\mathrm{h},i} ,
\end{align}
where $F_{\mathrm{h},i}$ is the total hydrodynamic force acting on the particle resulting from the pressure and viscous stresses at the particle surface,$V_{\mathrm{p},q}$ is the volume of the particle, and $\rho_{\mathrm{p},q}$ is the density of the particle. Note that gravity and other body force are not considered, although these do not change the analysis from a fundamental point of view. The hydrodynamic force is obtained by integrating the fluid stresses over the particle surface
\begin{align}
    F_{\mathrm{h},i} = \int\displaylimits_{\partial\Omega_{\mathrm{p},q}}\left(-p \delta_{ij} + \mu_{\mathrm{f}} \left(\dfrac{\partial u_i}{\partial x_j}+\dfrac{\partial u_j}{\partial x_i}\right)\right)\hat{n}_j\mathrm{d}A,
\end{align}
where $\delta_{ij}$ is the Kronecker delta. We assume that the particle surface normal vector $\hat{n}_j$ points towards the outside of the particle. \\
From a computational point of view, it is desirable to consistently smooth the high gradients arising near the particle to allow for a coarser resolution of the flow. The smoothing can be realized by volume-filtering the equations governing the flow using a filter kernel $g$, as originally proposed by \citet{Anderson1967} and later extended to the Euler-Lagrange framework by \citet{Capecelatro2013}. An arbitrary quantity $\varPhi$ that is defined in the fluid domain $\Omega_{\mathrm{f}}$ can be volume-filtered according to
\begin{align}
\label{eq:volumefiltering}
    \epsilon_{\mathrm{f}}(\boldsymbol{x}) \overline{\varPhi}(\boldsymbol{x}) = \int\displaylimits_{\Omega} I_{\mathrm{f}}(\boldsymbol{y}) \varPhi(\boldsymbol{y})g(|\boldsymbol{x}-\boldsymbol{y}|)\mathrm{d}V_y , 
\end{align}
where $\overline{\varPhi}(\boldsymbol{x})$ is the filtered quantity, $\Omega=\Omega_{\mathrm{f}}\cup \Omega_{\mathrm{p}}$ is the union of the fluid and the particle domain, and $\epsilon_{\mathrm{f}}(\boldsymbol{x})$ is the fluid volume fraction. The fluid indicator function is defined as
\begin{align}
    I_\mathrm{f}(\boldsymbol{x})=\begin{cases}
        1 & \text{if } \boldsymbol{x} \in \Omega_\mathrm{f}\\
        0 & \text{else }.
    \end{cases}
\end{align}
A valid filter kernel has to satisfy 
\begin{align}
   \int\displaylimits_{\Omega}g(|\boldsymbol{x}|) \mathrm{d}V_x = 1.
\end{align}
The kernel possesses a characteristic width $\delta$ that we refer to as the filter width. Applying the volume-filtering operation to the NSE gives (see \citet{Anderson1967})
\begin{align}
\label{eq:unclosedcontinuity}
    \dfrac{\partial \epsilon_{\mathrm{f}}}{ \partial t} + \dfrac{\partial}{\partial x_i}(\epsilon_{\mathrm{f}}\Bar{u}_i) &= 0, \\
    \label{eq:unclosedmomentum}
    \rho_{\mathrm{f}}\dfrac{\partial \epsilon_{\mathrm{f}}\Bar{u}_i}{\partial t} + \rho_{\mathrm{f}}\dfrac{\partial}{\partial x_j}(\epsilon_{\mathrm{f}}\overline{u_i u_j}) &= -\dfrac{\partial \epsilon_{\mathrm{f}}\Bar{p}}{\partial x_i}+\mu_{\mathrm{f}}\dfrac{\partial}{\partial x_j}\left[ \epsilon_{\mathrm{f}}\left( \overline{\dfrac{\partial u_i}{\partial x_j}} + \overline{\dfrac{\partial u_j}{\partial x_i}} \right) \right] - s_i
\end{align}
with the momentum source term expressed as a sum over all particles $q$
\begin{align}
    s_i = \sum_q \int\displaylimits_{\partial\Omega_{\mathrm{p},q}}g(|\boldsymbol{x}-\boldsymbol{y}|)\left(-p \delta_{ij} + \mu_{\mathrm{f}} \left(\dfrac{\partial u_i}{\partial y_j}+\dfrac{\partial u_j}{\partial y_i}\right)\right)\hat{n}_j\mathrm{d}A_y,
\end{align}
where we dropped the spatial dependence for simplicity. Note that volume-filtered quantities are fundamentally different than their respective unfiltered relatives. A kinetic energy derived from the volume-filtered velocity, e.g., $K_{\mathrm{VF}}=\epsilon_{\mathrm{f}}\Bar{u}_i\Bar{u}_i/2$ does not necessarily obey the same conservation laws as $K=u_i u_i/2$. The volume-filtered scales only represent a portion of the total scales of interest and kinetic energy may be exchanged between the flow present at the volume-filtered scales and the flow present at the smaller subfilter scales. A direct projection of the physical principles to the volume-filtered quantities can lead to wrong conclusions about the consistency of numerical methods for particle-laden flows. \\
Equations \eqref{eq:unclosedcontinuity} and \eqref{eq:unclosedmomentum} contain quantities that are not filtered, namely the fluid velocity and pressure in the particle momentum source term $s_i$, the fluid velocity in the advective term, and the fluid velocity gradient in the viscous term. Therefore, a general solution of these governing equations is not possible without providing closures for the unfiltered terms. The advective term, the second term on the left-hand side of equation \eqref{eq:unclosedmomentum}, is typically replaced with 
\begin{align}
    \dfrac{\partial}{\partial x_j}(\epsilon_{\mathrm{f}}\Bar{u}_i \Bar{u}_j), \nonumber
\end{align}
which we refer to as single volume fraction advective term, as it is commonly used in volume-filtered Euler-Lagrange simulations (see, e.g., \citet{Capecelatro2013,Ireland2017,Mallouppas2013b,Arolla2015}). This replacement of the advective term is only valid if the single volume fraction advective term is subtracted and the original advective term of equation \eqref{eq:unclosedmomentum} is added, which can be realized by introducing a subfilter stress tensor, $\tau_{\mathrm{sfs},ij}$. Therefore, the subfilter stress tensor accounts for volume-filtering the fluid velocity instead of its dyadic product in the advective term. Since the solution of volume filtered momentum equation provides an expression for $\epsilon_{\mathrm{f}}\Bar{u}_i$ and $\epsilon_{\mathrm{f}}\Bar{p}$, a division by $\epsilon_{\mathrm{f}}$ is required to compute the single volume fraction advective term. In cases where the fluid volume fraction is locally very small, small errors of the numerical solution of the volume-filtered momentum equation may lead to instabilities as it is reported by \citet{Dave2023}. A small fluid volume fraction arises for small filter width to particle size ratios in the center of the particle or even for larger filter width to particle size ratios in the case of many closely packed particles. Numerical interventions such as an artificial increase of the filter width are required in these cases \citep{Link2005,Jing2016,Chen2022}. Using the identity that the sum of the fluid volume fraction and the particle volume fractions $\epsilon_{\mathrm{p},q}$ for every particle $q$ equals one,
\begin{align}
\label{eq:conservationvolumefraction}
    \epsilon_{\mathrm{f}} +\sum\displaylimits_q \epsilon_{\mathrm{p},q} = 1,
\end{align}
we can decompose the single volume fraction advective term as follows:
\begin{align}
\label{eq:identityadvectiveterm}
     \dfrac{\partial \epsilon_{\mathrm{f}}\Bar{u}_i\Bar{u}_j}{\partial x_j} = \dfrac{\partial \epsilon_{\mathrm{f}} \Bar{u}_i \epsilon_{\mathrm{f}} \Bar{u}_j}{\partial x_j} + \sum_q \dfrac{\partial \epsilon_{\mathrm{f}}\Bar{u}_i \epsilon_{\mathrm{p},q} \Bar{u}_j}{\partial x_j}.
\end{align}
The first term on the right-hand side turns out to be numerically much more suitable as advective term in the volume-filtered NSE than the single volume fraction advective term, since no division by the (potentially very small) fluid volume fraction is required to obtain the filtered fluid flow velocity. We can rewrite the volume-filtered momentum equation \eqref{eq:unclosedmomentum} such that terms that are closed and terms that require closure are separated:
\begin{align}
\label{eq:momentumwithclosures}
    \rho_{\mathrm{f}}\dfrac{\partial \epsilon_{\mathrm{f}}\Bar{u}_i}{\partial t} + \rho_{\mathrm{f}}\dfrac{\partial}{\partial x_j}(\epsilon_{\mathrm{f}}\Bar{u}_i\epsilon_{\mathrm{f}}\Bar{u}_j) = -\dfrac{\partial \epsilon_{\mathrm{f}}\Bar{p}}{\partial x_i}+\mu_{\mathrm{f}} \dfrac{\partial^2\epsilon_{\mathrm{f}}\Bar{u}_i}{\partial x_j \partial x_j} - \sum_q s_{q,i}+ \mu_{\mathrm{f}} \mathcal{E}_i \nonumber \\ -\rho_{\mathrm{f}}\dfrac{\partial }{\partial x_j}\tau_{\mathrm{sfs},ij},
\end{align}
where $\mathcal{E}_i$ is the closure originating from switching the filter and spatial derivative in the viscous term
\begin{align}
\label{eq:definitionviscousclosure}
    \mathcal{E}_i = \dfrac{\partial}{\partial x_j}\left[ \epsilon_{\mathrm{f}}\left( \overline{\dfrac{\partial u_i}{\partial x_j}} + \overline{\dfrac{\partial u_j}{\partial x_i}} \right) \right] - \dfrac{\partial^2\epsilon_{\mathrm{f}}\Bar{u}_i}{\partial x_j \partial x_j},
\end{align}
and $\tau_{\mathrm{sfs},ij}$ is the subfilter stress tensor, which arises from the advective term
\begin{align}
\label{eq:definitionsubfilterstress}
    \tau_{\mathrm{sfs},ij} = \epsilon_{\mathrm{f}} \overline{u_i u_j} - \epsilon_{\mathrm{f}}\Bar{u}_i \epsilon_{\mathrm{f}}\Bar{u}_j =  \epsilon_{\mathrm{f}} \overline{u_i u_j} - \epsilon_{\mathrm{f}}\Bar{u}_i\Bar{u}_j + \sum_q\epsilon_{\mathrm{f}}\Bar{u}_i \epsilon_{\mathrm{p},q}\Bar{u}_j,
\end{align}
which can be written as a function of $\epsilon_{\mathrm{f}}\Bar{u}_i \epsilon_{\mathrm{f}}\Bar{u}_j$, which we refer to as double volume fraction advective term, or the single volume fraction advective term $\epsilon_{\mathrm{f}}\Bar{u}_i\Bar{u}_j$. Furthermore, $s_{q,i}$ is the particle momentum source originating from the particle with the index $q$ and given as
\begin{align}
\label{eq:particlemomentumsourceq}
   s_{q,i}= \int\displaylimits_{\partial\Omega_{\mathrm{p},q}}g(|\boldsymbol{x}-\boldsymbol{y}|)\left(-p \delta_{i j} + \mu_{\mathrm{f}} \left(\dfrac{\partial u_{i}}{\partial y_j}+\dfrac{\partial u_j}{\partial y_{i}}\right)\right)\hat{n}_j\mathrm{d}A_y.
\end{align}
Note that equation \eqref{eq:particlemomentumsourceq} depends on unfiltered quantities. The total hydrodynamic force acting on each particle, $|\boldsymbol{F}_{\mathrm{h},q}|=|\int_{\Omega}\boldsymbol{s}_q\mathrm{d}V|$, is typically modeled using empirical correlations. What remains to be closed is the distribution of the hydrodynamic force in space, which is expressed by the regularization kernel $\mathcal{K}_q$. In the direction of the resultant hydrodynamic force, which is indicated by the index $\alpha$, the regularization kernel is defined as
\begin{align}
s_{q,\alpha}=\mathcal{K}_q(|\boldsymbol{x}-\boldsymbol{x}_{\mathrm{p},q}|)F_{\mathrm{h},q,\alpha},
\end{align}
where $\boldsymbol{x}_{\mathrm{p},q}$ is the position of the particle center of the particle with the index $q$. Note that the particle momentum source may be non-zero in other directions than the direction of the resultant force, although its spatial integral is zero. The regularization kernel $\mathcal{K}_{q}$ is influenced by, e.g., the Reynolds number, velocity gradients, other particles or walls in the environment of the particle.\\
Note that the second term on the right-hand side of equation \eqref{eq:identityadvectiveterm} is absorbed in the subfilter stress tensor, such that equation \eqref{eq:momentumwithclosures} is equivalent to equation \eqref{eq:unclosedmomentum}. The double volume fraction advective term in combination with providing the closures in equation \eqref{eq:momentumwithclosures} allows to use a single-phase flow solver if the solution variables are $\epsilon_{\mathrm{f}}\Bar{u}_i$ and $\epsilon_{\mathrm{f}}\Bar{p}$. \\
Although the fluid viscosity is assumed to be constant to avoid additional unclosed terms, an additional spatially varying viscosity may be added after volume-filtering the fluid momentum equation, e.g., as part of the LES subgrid-scale modeling, resulting in a spatially varying total viscosity, $\mu_{\mathrm{tot}}$. In this case, the viscous term in the volume-filtered momentum equation \eqref{eq:momentumwithclosures} has the following form
\begin{align}
    \dfrac{\partial}{\partial x_j}\left[ \mu_{\mathrm{tot}} \left( \dfrac{\partial \epsilon_{\mathrm{f}}\Bar{u}_i}{\partial x_j} +  \dfrac{\partial \epsilon_{\mathrm{f}}\Bar{u}_j}{\partial x_i}\right)  \right], \nonumber
\end{align}
which yields an alternative viscous closure, $\mathcal{E}_i^{\mathrm{alt}}$,
\begin{align}
    \mu_{\mathrm{f}}\mathcal{E}_i^{\mathrm{alt}} = \dfrac{\partial}{\partial x_j}\left[ \mu_{\mathrm{tot}}\epsilon_{\mathrm{f}}\left( \overline{\dfrac{\partial u_i}{\partial x_j}} + \overline{\dfrac{\partial u_j}{\partial x_i}} \right) \right] - \dfrac{\partial}{\partial x_j}\left[ \mu_{\mathrm{tot}} \left( \dfrac{\partial \epsilon_{\mathrm{f}}\Bar{u}_i}{\partial x_j} +  \dfrac{\partial \epsilon_{\mathrm{f}}\Bar{u}_j}{\partial x_i}\right)  \right].
\end{align}

\subsection{Galilean invariance}
\label{ssec:galileaninvariance}
Specific care has to be taken to ensure Galilean invariance of the momentum equation when closures are derived, which is much less trivial than in the single-phase flow case. Galilean invariance is satisfied if the governing equations are invariant with respect to a constant velocity of the frame of reference $ \boldsymbol{u}_{\mathrm{ref}}$, such that
\begin{align}
    \mathcal{U}_i(\boldsymbol{x},t) = u_i(\boldsymbol{x} - \boldsymbol{u}_{\mathrm{ref}}t,t) +  u_{\mathrm{ref},i}, \\
    \mathcal{V}_i(\boldsymbol{x},t) = v_i(\boldsymbol{x} - \boldsymbol{u}_{\mathrm{ref}}t,t) +  u_{\mathrm{ref},i},
\end{align}
where $\mathcal{U}_i$ and $\mathcal{V}_i$ are the fluid and particle velocity in the fixed frame of reference and $u_i$ and $v_i$ are the fluid and particle velocity in the frame of reference moving with the velocity $\boldsymbol{u}_{\mathrm{ref}}$. It can be easily verified that the right-hand side of the volume-filtered momentum equation \eqref{eq:unclosedmomentum} satisfies the Galilean invariance. For the left-hand side we find
\begin{align}
\label{eq:Galileaninvariancetemporalderivative}
    \dfrac{\partial \epsilon_{\mathrm{f}} \Bar{\mathcal{U}}_i}{\partial t} = \dfrac{\partial \epsilon_{\mathrm{f}} \Bar{u}_i}{\partial t} - u_{\mathrm{ref},j}\dfrac{\partial \epsilon_{\mathrm{f}}\Bar{u}_i}{\partial x_j} - u_{\mathrm{ref},i}\dfrac{\partial \epsilon_{\mathrm{f}}\Bar{u}_j}{\partial x_j} - u_{\mathrm{ref},i}u_{\mathrm{ref},j}\dfrac{\partial \epsilon_{\mathrm{f}}}{\partial x_j},
\end{align}
using the continuity equation \eqref{eq:unclosedcontinuity}, and 
\begin{align}
    \dfrac{\partial \epsilon_{\mathrm{f}}\overline{\mathcal{U}_i \mathcal{U}_j}}{\partial x_j} = \dfrac{\partial \epsilon_{\mathrm{f}}\overline{u_i u_j}}{\partial x_j} + u_{\mathrm{ref},j}\dfrac{\partial \epsilon_{\mathrm{f}}\Bar{u}_i}{\partial x_j} + u_{\mathrm{ref},i}\dfrac{\partial \epsilon_{\mathrm{f}}\Bar{u}_j}{\partial x_j} + u_{\mathrm{ref},i}u_{\mathrm{ref},j}\dfrac{\partial \epsilon_{\mathrm{f}}}{\partial x_j}.
\end{align}
The sum of both terms leads to the original left-hand side of equation \eqref{eq:unclosedmomentum}, which verifies the Galilean invariance. In the same manner, it can be shown that if we replace the actual advective term $\dfrac{\partial}{\partial x_j}(\epsilon_{\mathrm{f}}\overline{u_i u_j})$ with the single volume fraction advective term $\dfrac{\partial}{\partial x_j}(\epsilon_{\mathrm{f}}\Bar{u}_i\Bar{u}_j)$, Galilean invariance of the momentum equations is achieved even if the subfilter stress tensor is not modeled and fully neglected. With the double volume fraction advective term $\dfrac{\partial}{\partial x_j}(\epsilon_{\mathrm{f}}\Bar{u}_i\epsilon_{\mathrm{f}}\Bar{u}_j)$, the subfilter stress tensor is not Galilean invariant. Consequently, at least one term that cancels out the terms induced by a change of frame of reference has to be included as part of the model for the subfilter stress tensor. \\
Changing the frame of reference of the double volume fraction advective term causes desired and undesired additional terms:
\begin{align} 
      \dfrac{\partial \epsilon_{\mathrm{f}} \Bar{u}_i \epsilon_{\mathrm{f}} \Bar{u}_j}{\partial x_j} =\underbrace{\dfrac{\partial \epsilon_{\mathrm{f}}\Bar{u}_i\Bar{u}_j}{\partial x_j}}_{\text{terms compensated by equation \eqref{eq:Galileaninvariancetemporalderivative}}}- \underbrace{\sum_q \dfrac{\partial \epsilon_{\mathrm{f}}\Bar{u}_i \epsilon_{\mathrm{p},q} \Bar{u}_j}{\partial x_j}}_{\text{additional terms}}.
\end{align}
The terms that arise by changing the frame of reference of the first term on the right-hand side are required to compensate the terms arising from the temporal velocity derivative equation \eqref{eq:Galileaninvariancetemporalderivative}. However, the second term on the right-hand side causes terms that are not compensated without further interventions, i.e., the volume-filtered momentum equation is not Galilean invariant. Therefore, a compensating term has to be added. The minimal expression for compensating the additional terms is
\begin{align}
    \tau_{\mathrm{sfs},ij}^{\mathrm{G}} = \sum_{q}\left[ \epsilon_{\mathrm{p},q} v_{q,j}\epsilon_{\mathrm{f}} \Bar{u}_i + \epsilon_{\mathrm{p},q} v_{q,i}\epsilon_{\mathrm{f}} \Bar{u}_j - \epsilon_{\mathrm{p},q} v_{q,i} \epsilon_{\mathrm{f}} v_{q,j}\right].
\end{align}
We consider $\tau_{\mathrm{sfs},ij}^{\mathrm{G}}$ as the Galilean invariance part of the subfilter stress tensor. This term does nothing else but compensate the additional terms arising from a change of the frame of reference. It is not a model for the physics in the particle fixed frame, but it ensures that the interscale energy and momentum transfer is the same in all frames of reference. To demonstrate the effect of $\tau_{\mathrm{sfs},ij}^{\mathrm{G}}$, consider the following subfilter stress tensor in the case of flow around a fixed particle
\begin{align}
    \tau_{\mathrm{sfs},ij} = \epsilon_{\mathrm{f}} \overline{u_i u_j} - \epsilon_{\mathrm{f}}\Bar{u}_i \epsilon_{\mathrm{f}}\Bar{u}_j.
\end{align}
Changing the frame of reference, such that the particle moves, leads to a different field of $\tau_{\mathrm{sfs},ij}$, while $\tau_{\mathrm{sfs},ij}+\tau_{\mathrm{sfs},ij}^{\mathrm{G}}$ is exactly the same for both frames of reference. \\
The behavior of the volume-filtered equations when the frame of reference is changed, can be intuitively understood when the definition of the volume-filtering equation \eqref{eq:volumefiltering} is considered. If the flow quantity $\varPhi$ in equation \eqref{eq:volumefiltering} has the dimension of a velocity, its value changes when the frame of reference is moved. Due to the indicator function, however, the value of the integrand is always zero outside of the fluid region. Consequently, gradients of flow quantities close to the interface are dependent on the frame of reference, which affects all terms in the continuity and momentum equation that contain spatial derivatives of volume-filtered velocities. \\
A main outcome of the Galilean invariance is that a temporal change of the volume fraction, caused by a change of the frame of reference or a moving particle, leads to additional terms if a term contains spatial gradients of $\epsilon_{\mathrm{f}}\Bar{u}_i$. This has to be considered in the closure modeling for the volume-filtered momentum equation \eqref{eq:momentumwithclosures}.

\section{Investigation of the closures}
\label{sec:closures}
In this section, we discuss the closure terms in the volume-filtered momentum equation \eqref{eq:momentumwithclosures}, derive analytical expressions or propose models for the closures, and provide a foundation for more advanced models.

\subsection{Viscous closure $\mathcal{E}_i$}
\label{ssec:viscousclosure}
The term $\mathcal{E}_i$ arises because the volume-filtered velocity gradients in the viscous stresses of equation \eqref{eq:unclosedmomentum} are replaced with the gradients of the volume-filtered velocity. Formally, we can write $\mathcal{E}_i$ as the difference between the volume-filtered viscous term and the Laplacian of $\epsilon_{\mathrm{f}}\Bar{u}_i$ (see equation \eqref{eq:definitionviscousclosure}). The effect of $\mathcal{E}_i$ is typically modeled by an additional viscosity \citep{Zhang1997a,Patankar2001a,Capecelatro2013} or neglected \citep{Subramaniam2022}. To the best of the authors knowledge, however, no study exists that proves the validity of either of the two assumptions. In the present work, we derive an expression for $\mathcal{E}_i$ analytically that is valid without any further assumptions. \\
We can approach the problem of finding an expression for $\mathcal{E}_i$ to search for a closure for $\epsilon_{\mathrm{f}}\overline{\dfrac{\partial u_i}{\partial x_j}}-\dfrac{\partial}{\partial x_j}(\epsilon_{\mathrm{f}} \overline{u}_i)$. With the definition of the volume-filtering equation \eqref{eq:volumefiltering} we obtain
\begin{align}
    \epsilon_{\mathrm{f}} \overline{\dfrac{\partial u_i}{\partial x_j}}-\dfrac{\partial}{\partial x_j}(\epsilon_{\mathrm{f}} \overline{u}_i) &= \int\displaylimits_{\Omega_f}g(|\boldsymbol{x-y}|)\dfrac{\partial u_i(\boldsymbol{y})}{\partial y_j} - \dfrac{\partial g(|\boldsymbol{x-y}|)}{\partial x_j}u_i(\boldsymbol{y})dV_y \nonumber \\
    &= \int\displaylimits_{\Omega_f}g(|\boldsymbol{x-y}|)\dfrac{\partial u_i(\boldsymbol{y})}{\partial y_j} + \dfrac{\partial g(|\boldsymbol{x-y}|)}{\partial y_j}u_i(\boldsymbol{y})dV_y \nonumber \\
    &= \int\displaylimits_{\Omega_f} \dfrac{\partial }{\partial y_j}\left( g(|\boldsymbol{x-y}|) u_i(\boldsymbol{y}) \right)dV_y.
\end{align}
For the following analysis, a distinction between the volume fraction of each particle is required if more than one particle is considered. We consider multiple particles by assigning a particle volume fraction $\epsilon_{\mathrm{p},q}$ to every particle such that equation \eqref{eq:conservationvolumefraction} is satisfied. By applying the divergence theorem, considering that the normal vector $n_j$ points towards the inside of the particle, and the fluid velocity equals the particle velocity at the particle surface, we can further simplify
\begin{align}
\label{eq:switichingfilteringderivative}
    \epsilon_{\mathrm{f}} \overline{\dfrac{\partial u_i}{\partial x_j}}-\dfrac{\partial}{\partial x_j}(\epsilon_{\mathrm{f}} \overline{u}_i) &=\int\displaylimits_{\partial \Omega_\mathrm{f}}g(|\boldsymbol{x-y}|) u_i(\boldsymbol{y}) n_j dA_y \nonumber \\
    &= -\sum\displaylimits_q\int\displaylimits_{\partial \Omega_{\mathrm{p},q}}g(|\boldsymbol{x-y}|) u_i(\boldsymbol{y}) \hat{n}_j dA_y \nonumber \\
    &= -\sum\displaylimits_q v_{q,i}\int\displaylimits_{\partial \Omega_{\mathrm{p},q}}g(|\boldsymbol{x-y}|)\hat{n}_j dA_y \nonumber \\
    &= \sum\displaylimits_q v_{q,i}\int\displaylimits_{\Omega_{\mathrm{p},q}}\dfrac{\partial g(|\boldsymbol{x-y}|)}{\partial x_j}dV_y \nonumber \\
    &= \sum\displaylimits_q v_{q,i}\dfrac{\partial \epsilon_{\mathrm{p},q}}{\partial x_j},
\end{align}
where $\hat{n}_j$ represents the normal vector pointing into the fluid. Although particle rotation is excluded in the present study, it is worth noting that a pure rotation of the spherical particle does not contribute to the closure (see appendix \ref{ap:rotationviscouclosure}).
From equation \eqref{eq:switichingfilteringderivative}, we can conclude that if the volume fraction does not change in time, the filtered velocity gradient and the gradient of the filtered velocity are equivalent. \\
Since the volume of each particle stays constant, we can write
\begin{align}
    \dfrac{\mathrm{D}\epsilon_{\mathrm{p},q}}{\mathrm{D}t} = 0 = \dfrac{\partial \epsilon_{\mathrm{p},q}}{\partial t} + v_{q,i}\dfrac{\partial \epsilon_{\mathrm{p},q}}{\partial x_i}.
\end{align}
Using the conservation of volume fraction, equation \eqref{eq:conservationvolumefraction}, we get
\begin{align}
    \dfrac{\partial \epsilon_{\mathrm{f}}}{\partial t} = \dfrac{\partial (1-\sum\displaylimits_q \epsilon_{\mathrm{p},q})}{\partial t} = \sum\displaylimits_q v_{q,i}\dfrac{\partial \epsilon_{\mathrm{p},q}}{\partial x_i},
\end{align}
and by incorporating the continuity equation \eqref{eq:unclosedcontinuity}
\begin{align}
\label{eq:divergencevolumefiltered}
    \dfrac{\partial \epsilon_{\mathrm{f}} \Bar{u}_i}{\partial x_i} = -\sum\displaylimits_q v_{q,i}\dfrac{\partial \epsilon_{\mathrm{p},q}}{\partial x_i}.
\end{align}
By inserting equation \eqref{eq:switichingfilteringderivative} into the definition of the viscous closure equation \eqref{eq:definitionviscousclosure}, we obtain
\begin{align}
\label{eq:viscousclosure}
    \mathcal{E}_i &= \dfrac{\partial}{\partial x_j}\left[  \dfrac{\partial \epsilon_{\mathrm{f}} \Bar{u}_i}{\partial x_j} + \sum\displaylimits_q v_{q,i}\dfrac{\partial \epsilon_{\mathrm{p},q}}{\partial x_j} + \dfrac{ \partial \epsilon_{\mathrm{f}} \Bar{u}_j}{\partial x_i} +\sum\displaylimits_q v_{q,j}\dfrac{\partial \epsilon_{\mathrm{p},q}}{\partial x_i}\right] - \dfrac{\partial^2\epsilon_{\mathrm{f}}\Bar{u}_i}{\partial x_j \partial x_j} \nonumber \\
    &= \dfrac{\partial}{\partial x_j}\left[ \sum\displaylimits_q v_{q,i}\dfrac{\partial \epsilon_{\mathrm{p},q}}{\partial x_j} + \dfrac{ \partial \epsilon_{\mathrm{f}} \Bar{u}_j}{\partial x_i} +\sum\displaylimits_q v_{q,j}\dfrac{\partial \epsilon_{\mathrm{p},q}}{\partial x_i}\right] \nonumber\\
    &= \sum\displaylimits_q v_{q,i}\dfrac{\partial^2 \epsilon_{\mathrm{p},q}}{\partial x_j \partial x_j} + \dfrac{\partial}{\partial x_i}\left( \dfrac{ \partial \epsilon_{\mathrm{f}} \Bar{u}_j}{\partial x_j} \right)+\sum\displaylimits_q v_{q,j}\dfrac{\partial^2 \epsilon_{\mathrm{p},q}}{\partial x_i \partial x_j} \nonumber\\
    &= \sum\displaylimits_q v_{q,i}\dfrac{\partial^2 \epsilon_{\mathrm{p},q}}{\partial x_j \partial x_j},
\end{align}
where we used equation \eqref{eq:divergencevolumefiltered}. In numerical simulations, the particle volume fraction is an easily accessible quantity with a known analytical expression for spherical particles that can be found, for instance, in \citet{Balachandar2022}, using a Gaussian filter kernel $g$. Its second spatial derivative can be computed numerically or analytically and $\mathcal{E}_i$ can be added as an explicit source term in the fluid momentum equation \eqref{eq:momentumwithclosures}. \\
Together with the closure $\mathcal{E}_i$, the viscous contribution of equation \eqref{eq:momentumwithclosures} is Galilean invariant. Consequently, the closure must be included to ensure a physical representation of the volume-filtered fluid-particle system. \\
If the total viscosity, $\mu_{\mathrm{tot}}$, varies in space because of an additional viscosity added to the already volume-filtered momentum equation, the alternative viscous term and the alternative viscous closure are applied, which is analogously obtained as
\begin{align}
\label{eq:alternativeviscousclosure}
    \mu_{\mathrm{f}}\mathcal{E}_i^{\mathrm{alt}} = \dfrac{\partial }{\partial x_j}\left[ \mu_{\mathrm{tot}} \sum_{q} \left( v_{q,i}\dfrac{\partial \epsilon_{\mathrm{p},q}}{\partial x_j} +v_{q,j}\dfrac{\partial \epsilon_{\mathrm{p},q}}{\partial x_i} \right) \right].
\end{align}

\subsection{Particle momentum source $s_i$}
\begin{figure}
    \centering
    \includegraphics[scale=0.75]{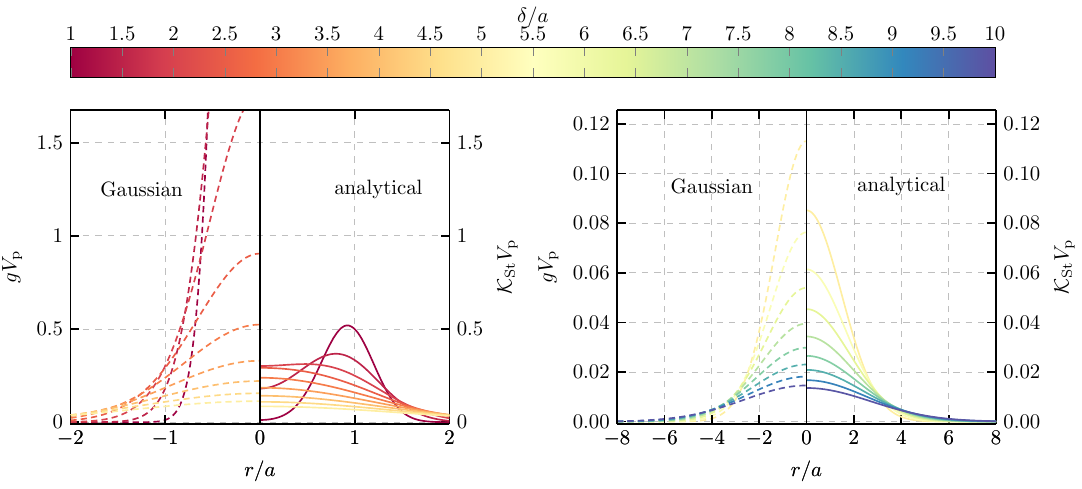}
    \caption{Gaussian kernel (dashed lines) and the analytical kernel in the Stokes flow (solid lines) for different filter widths. The color indicates the different filter widths. Left for $1\leq \delta/a \leq 5$ and right for $5 \leq \delta/a \leq 10$.}
    \label{fig:kernel}
\end{figure}

The particle momentum source term, $s_i$, represents the momentum exchange between the particles and the filtered flow scales. We exclude the discussion of the modeling of the hydrodynamic force on the particle, $F_{\mathrm{h},i}$. Instead, we discuss the regularization kernel, i.e., how $s_i$ varies in space, which is typically approximated with a Gaussian, a polynomial approximation of a Gaussian or a top-hat function \citep{Crowe1977,Maxey2017,Maxey2001,Evrard2020a,Poustis2019,Keane2023}. It is commonly argued that if the filter width is much larger than the size of the particle, i.e., $\delta\gg a$, the filter kernel $g$ varies insignificantly across the volume of the particle and can be treated as a constant for the integration \citep{Capecelatro2013}: 
\begin{align}
    s_i &= \sum_q \int\displaylimits_{\partial\Omega_{\mathrm{p},q}}g(|\boldsymbol{x}-\boldsymbol{y}|)\left(-p \delta_{ij} + \mu_{\mathrm{f}} \left(\dfrac{\partial u_i}{\partial y_j}+\dfrac{\partial u_j}{\partial y_i}\right)\right)n_j\mathrm{d}A_y  \nonumber \\
    &\approx \sum_q F_{\mathrm{h},q,i}g(|\boldsymbol{x}-\boldsymbol{x}_{\mathrm{p},q}|),
\end{align}
where $\boldsymbol{x}_{\mathrm{p},q}$ is the center of the respective particle. It is not clear, however, at which filter width this is a valid assumption and what the effect of this approximation is on the velocity field. This is discussed in more detail in section \ref{sec:evaluationclosures}. \\
For a general expression for $s_i$, the exact pressure and velocity field at the surface of the particle is required and its convolution with the filter kernel $g$ must be integrable. We find such an expression in the limit of $\mathrm{Re}\rightarrow 0$, where the flow is governed by the Stokes flow equations. By exploiting that the fluid stresses in Stokes flow are constant over the whole spherical particle surface with radius $a$, we get 
\begin{align}
\label{eq:approximationfors}
    s_i &= \int\displaylimits_{\partial\Omega_{\mathrm{p}}}g(|\boldsymbol{x}-\boldsymbol{y}|)(-p \delta_{ij} + \mu_{\mathrm{f}} (\dfrac{\partial u_i}{\partial y_j}+\dfrac{\partial u_j}{\partial y_i}))n_j\mathrm{d}A_y  \nonumber \\
    &= \dfrac{3\mu_{\mathrm{f}}(u_{\infty,i}-v_{i})}{2a}\int\displaylimits_{\partial\Omega_{\mathrm{p}}}g(|\boldsymbol{x}-\boldsymbol{y}|)\mathrm{d}A_y \nonumber \\
    &=  F_{\mathrm{h},i} \mathcal{K}_{\mathrm{St}}(|\boldsymbol{x}-\boldsymbol{x}_{\mathrm{p}}|),
\end{align}
where $u_{\infty,i}$ is the fluid velocity far away from the particle. If $g$ is assumed to be Gaussian, 
\begin{align}
    g(\boldsymbol{x}) = \dfrac{1}{(2\pi \sigma^2)^{3/2}}\exp\left( -\dfrac{|\boldsymbol{x}|^2}{2 \sigma^2} \right),
\end{align}
the explicit evaluation of the integral in equation \eqref{eq:approximationfors} gives the regularization kernel for an isolated spherical particle in Stokes flow
\begin{align}
    \mathcal{K}_{\mathrm{St}}(r) = \dfrac{\exp{\left(-\dfrac{(a+r)^2}{2\sigma^2}\right)}}{4\pi\sqrt{2 \pi \sigma^2}ra} \left( \exp{\left(\dfrac{2ar}{\sigma^2}\right)} - 1\right),
\end{align}
where $\sigma$ is the standard deviation of the Gaussian. For the remainder of this work, we assume the relation $\sigma = \delta \sqrt{2/9\pi}$. This somewhat arbitrary choice of the filter width is equal to the support of the Wendland kernel, a practical relevant polynomial approximation of a Gaussian, which is used, e.g., by \citet{Evrard2020a} to regularize the particle momentum source. \\
Note that an expression for the regularization kernel is not limited to a Gaussian filter kernel, but can be obtained for every filter kernel that can be integrated according to equation \eqref{eq:approximationfors}. It has to be highlighted, that the kernel $\mathcal{K}_{\mathrm{St}}$ is only valid in uniform Stokes flow over a single sphere in an infinite domain. At higher Reynolds number, in shear flow, in the vicinity of a wall, or if multiple particles are present, the particle momentum source is distributed differently in space. Note that although the Stokes flow equations are linear, that does not allow for a superposition of the kernels $\mathcal{K}_{\mathrm{St}}$ for multiple spheres, as the fluid stresses are no longer constant over the particle surface. \\
To assess if the commonly applied Gaussian kernel is a suitable choice for the regularization of the momentum source, it is compared to the analytical regularization kernel in Stokes flow. In figure \ref{fig:kernel}, the analytical regularization kernel of Stokes flow around a sphere $\mathcal{K}_{\mathrm{St}}$ is shown for different filter width to particle size ratios along with the Gaussian kernel $g$ that is commonly used to regularize the force in point-particle simulations. The Gaussian and the analytical kernel are normalized with the volume of the spherical particle, $V_{\mathrm{p}}=4\pi a^3/3$. The largest deviations are observed for small filter widths, where the Gaussian has much larger values than the analytical kernel. An interesting property of the analytical kernel is that for small filter widths it is centered at the particle surface, whereas the Gaussian kernel is centered at the particle center. As the filter width increases, the magnitudes and shapes of the two kernels approach each other and at filter widths of approximately $\delta/a\ge8$ only marginal differences between the kernels can be observed. For smaller filter widths, however, significant deviations are observed. \\
As the fluid stresses become nonuniformly distributed over the particle surface because of a higher Reynolds number or the presence of other particles, the kernel looses its spherical symmetry. In such a case, $\mathcal{K}_{\mathrm{St}}$ becomes an approximation. The Gaussian kernel, however, is an approximation for all $\mathrm{Re}$. As shown in figure \ref{fig:kernel}, the Gaussian kernel is a very poor approximation of the real regularization kernel at small $\mathrm{Re}$ and small filter width to particle size ratios.

\subsection{Subfilter stress tensor $\tau_{\mathrm{sfs},ij}$}
\label{sec:modeltau}
The subfilter stress tensor, $\tau_{\mathrm{sfs},ij}$, originates from the difference of the volume-filtered tensor product of the fluid velocity vectors and the tensor product of the volume-filtered fluid velocity vectors. With the choice of the advective term in the volume-filtered fluid momentum equation \eqref{eq:unclosedmomentum}, a modeled subfilter stress tensor needs to at least contain a contribution to satisfy Galilean invariance $\tau_{\mathrm{sfs}, ij}^{\mathrm{G}}$, which is discussed in section \ref{ssec:galileaninvariance}. Furthermore, the non-linear subfilter interactions are modeled with $\tau_{\mathrm{sfs}, ij}^{\mathrm{inter}}$, which is Galilean invariant itself, such that
\begin{align}
    \tau_{\mathrm{sfs}, ij}^{\mathrm{mod}} = \tau_{\mathrm{sfs}, ij}^{\mathrm{G}} + \tau_{\mathrm{sfs}, ij}^{\mathrm{inter}}.
\end{align}
While the expression to compensate for additional terms arising from the change of the frame of reference $\tau_{\mathrm{sfs}, ij}^{\mathrm{G}}$ is known, the non-linear subfilter interactions $\tau_{\mathrm{sfs}, ij}^{\mathrm{inter}}$ require modeling. \\ 
A major difficulty for understanding and modeling the subfilter stress tensor is that, unlike other closures, it can be non-zero also in the absence of particles. When a turbulent single-phase flow with a filter width larger than the Kolmogorov length scale is considered, as is the case in classical LES, $\tau_{\mathrm{sfs},ij}$ accounts for the mainly dissipative effect of the unresolved turbulent velocity on the filtered scales. The effect of the subfilter stresses on the filtered scales is relatively well understood in single-phase flow cases such as homogeneous isotropic turbulence (see, e.g., \citet{Sagaut2005}). However, if particles are present, one does not even fully understand the effect of a single particle subject to laminar flow on the subfilter stress tensor, and especially not the effects of an additional turbulent background flow or of fluid velocity disturbances due to neighboring particles. \citet{Mehrabadi2015} studied the behavior of the pseudo-turbulent Reynolds stresses, $R_{ij}$, with different configurations of particle resolved direct numerical simulations. The pseudo-turbulent Reynolds stresses, $R_{ij}$, are defined as
\begin{align}
    R_{ij}  = \epsilon_{\mathrm{f}} \overline{u_i^{\prime} u_j^{\prime}},
\end{align}
where the subfilter velocity is given by
\begin{align}
    u_i^{\prime} = u_i - \Bar{u}_i.
\end{align}
Note that \citet{Mehrabadi2015} define the pseudo-turbulent Reynolds stresses based on ensemble averaging, which is related, but not equivalent to the volume-filtering studied in the present work. There have been attempts at modeling the trace of $R_{ij}$, the pseudo-turbulent kinetic energy, and reconstruct $R_{ij}$ based on it \citep{Mehrabadi2015,Shallcross2020}. However, finding a model for the pseudo-turbulent Reynolds stresses in the context of volume-filtering is not equivalent to finding a closure for the subfilter stress tensor. The first term of $\tau_{\mathrm{sfs},ij}$ can be decomposed as follows
\begin{align}
    \epsilon_{\mathrm{f}}\overline{u_i u_j} = \epsilon_{\mathrm{f}}\overline{ \Bar{u}_i \Bar{u}_j} + \epsilon_{\mathrm{f}}\overline{ \Bar{u}_i u^{\prime}_j}+ \epsilon_{\mathrm{f}}\overline{ u^{\prime}_i\Bar{u}_j } + R_{ij}.
\end{align}
Consequently, in order to model $\tau_{\mathrm{sfs},ij}$, a model for $R_{ij}$ is not sufficient. Although the importance of the remaining three terms is unknown for a particle-laden flow, it is known from single-phase turbulence that the contribution of these three terms is not negligible and, in some configurations, even dominate the contribution of $R_{ij}$ \citep{Ghate2020b,Laval2001,Hunt1990,Mann1994}. \\
There have been suggestions to model the subfilter stress tensor with an additional viscosity according to the Boussinesq hypothesis in turbulence \citep{Subramaniam2022,Capecelatro2013}. \citet{Hausmann2023} derived an eddy viscosity model for LES of particle-laden turbulent flows that accounts for the turbulence modulation by the particles, but with the assumption that the effect of the sub-Kolmogorov particles on the subfilter stress tensor is negligible if the turbulence is fully resolved. \\
To indicate the filter width of a volume-filtered flow quantity weighted with the fluid indicator function, $I_{\mathrm{f}}\varPhi$, we introduce the following notation
\begin{align}
\label{eq:volumefilteringprecise}
    \overline{I_{\mathrm{f}}\Phi}\vert_{\sigma} = \int\displaylimits_{\Omega} I_{\mathrm{f}}(\boldsymbol{y}) \varPhi(\boldsymbol{y})g(|\boldsymbol{x}-\boldsymbol{y}|)\mathrm{d}V_y, 
\end{align}
where $(.)\vert_{\sigma}$ indicates that the flow quantity is filtered using a Gaussian with a standard deviation $\sigma$. Assuming that the filtering kernel $g$ is Gaussian, it can be verified that the volume-filtering operation with equation \eqref{eq:volumefiltering} is equivalent to solving the diffusion equation for a flow quantity weighted with the fluid indicator function, $I_{\mathrm{f}}\varPhi$, (see, e.g., \citet{Johnson2021})
\begin{align}
\label{eq:differentialfiltering}
    \dfrac{\partial \overline{I_{\mathrm{f}}\Phi}\vert_{\sigma}}{\partial (\sigma^2)} = \dfrac{1}{2}\nabla^2\overline{I_{\mathrm{f}}\Phi}\vert_{\sigma},
\end{align}
with the initial condition 
\begin{align}
    \overline{I_{\mathrm{f}}\Phi}\vert_{\sigma=0} = I_{\mathrm{f}}\Phi.
\end{align}
The filtered result is
\begin{align}
    \overline{I_{\mathrm{f}}\Phi}\vert_{\sigma} = \epsilon_{\mathrm{f}} \overline{\Phi}.
\end{align}
With the definition of the non-linear advective closure equation \eqref{eq:definitionsubfilterstress}, we can find the exact relation (see appendix \ref{ap:subfilterstressequation})
\begin{align}
\label{eq:differentialfilteringtau}
    \dfrac{\partial \tau_{\mathrm{sfs},ij}\vert_{\sigma}}{\partial (\sigma^2)} = \dfrac{1}{2}\nabla^2\tau_{\mathrm{sfs},ij}\vert_{\sigma} + \dfrac{\partial \overline{I_{\mathrm{f}}u_i}\vert_{\sigma}}{\partial x_k}\dfrac{\partial \overline{I_{\mathrm{f}}u_j}\vert_{\sigma}}{\partial x_k},
\end{align}
with the initial condition 
\begin{align}
    \tau_{\mathrm{sfs},ij}\vert_{\sigma=0} = 0.
\end{align}
Equation \eqref{eq:differentialfilteringtau} is analog to the expression obtained from \citet{Johnson2021} for the subfilter stress tensor in a single-phase flow. Since the Green's function $G$ of equation \eqref{eq:differentialfilteringtau} is given as
\begin{align}
    G\vert_{\sigma}(|\boldsymbol{x}|) = \begin{cases} 
      g\vert_{\sigma}(|\boldsymbol{x}|) & \sigma^2\geq 0 \\
      0 & \sigma^2<0,
   \end{cases}
\end{align}
where $g\vert_{\sigma}$ is the Gaussian of standard deviation $\sigma$, its formal solution reads as the convolution integral 
\begin{align}
    \tau_{\mathrm{sfs},ij}\vert_{\sigma}(\boldsymbol{x}) = \int\displaylimits_0^{\sigma^2} \int\displaylimits_{\Omega} G\vert_{\sqrt{\sigma^2-\theta}}(|\boldsymbol{x}-\boldsymbol{s}|)\dfrac{\partial \overline{I_{\mathrm{f}}u_i(\boldsymbol{s})}\vert_{\sqrt{\theta}}}{\partial x_k}\dfrac{\partial \overline{I_{\mathrm{f}}u_j(\boldsymbol{s})}\vert_{\sqrt{\theta}}}{\partial x_k}\mathrm{d}\boldsymbol{s}\mathrm{d}\theta.
\end{align}
After evaluating the inner integral, which is essentially another filtering operation, we obtain
\begin{align}
\label{eq:formalsolutiontau}
    \tau_{\mathrm{sfs},ij}\vert_{\sigma} &= \int\displaylimits_0^{\sigma^2} \left.\overline{ \dfrac{\partial\overline{ I_{\mathrm{f}}u_i}\vert_{\sqrt{\theta}}}{\partial x_k} \dfrac{\partial\overline{ I_{\mathrm{f}}u_j}\vert_{\sqrt{\theta}}}{\partial x_k}}\right\vert_{\sqrt{\sigma^2-\theta}}\mathrm{d}\theta \nonumber \\ 
    &= \sigma^2 \left.\overline{I_{\mathrm{f}}\dfrac{\partial u_i}{\partial x_k}}\right\vert_{\sigma}\left.\overline{I_{\mathrm{f}}\dfrac{\partial u_j}{\partial x_k}}\right\vert_{\sigma} + \int\displaylimits_0^{\sigma^2} \left.\overline{ \dfrac{\partial\overline{ I_{\mathrm{f}}u_i}\vert_{\sqrt{\theta}}}{\partial x_k} \dfrac{\partial\overline{ I_{\mathrm{f}}u_j}\vert_{\sqrt{\theta}}}{\partial x_k}}\right\vert_{\sqrt{\sigma^2-\theta}}\mathrm{d}\theta  \nonumber\\  &-\int\displaylimits_0^{\sigma^2} \left.\overline{ \left.\overline{I_{\mathrm{f}}\dfrac{\partial u_i}{\partial x_k}}\right\vert_{\sqrt{\theta}} }\right\vert_{\sqrt{\sigma^2-\theta}} \left.\overline{ \left.\overline{I_{\mathrm{f}}\dfrac{\partial u_j}{\partial x_k}}\right\vert_{\sqrt{\theta}} }\right\vert_{\sqrt{\sigma^2-\theta}}\mathrm{d}\theta,
\end{align}
where we use the relation (see, e.g., \citet{Johnson2021})
\begin{align}
\overline{\overline{I_{\mathrm{f}}u}_i\vert_{\sigma_1}}\vert_{\sigma_2} = \overline{I_{\mathrm{f}}u}_i\vert_{\sqrt{\sigma_1^2 + \sigma_2^2}}.
\end{align}
The first term on the right-hand side of equation \eqref{eq:formalsolutiontau} contains only filtered quantities, whereas the remaining terms also consists of the fluid velocity at scales that are smaller than $\sigma$. The first and the last term on the right-hand side of equation \eqref{eq:formalsolutiontau} are Galilean invariant, but the second term is not. The subfilter stress tensor, as defined in equation \eqref{eq:definitionsubfilterstress}, is not Galilean invariant. For single-phase turbulence, the contribution of the integral terms to the kinetic energy transfer are minor compared to the contribution of the first term for small filter widths. The first term can be computed from knowledge of volume-filtered quantities only, which essentially leads to a well known LES model for single-phase flow turbulence, sometimes referred to as the non-linear model \citep{Liu1994b,Borue1998}. This model was first derived by \citet{Leonard1975} with a Taylor expansion of the filtered velocity in space. It has been shown by \citet{Johnson2021} that the resolved contribution of equation \eqref{eq:formalsolutiontau} is dominant for small filter widths in single-phase flow turbulence. \citet{Borue1998} showed that the non-linear model strongly correlates with the subfilter stress tensor. Therefore, the subfilter stress tensor is proposed to be modeled as the resolved contribution of equation \eqref{eq:formalsolutiontau}, i.e., the first term on the right-hand side of equation \eqref{eq:formalsolutiontau}
\begin{align}
\label{eq:nonlinearmodel}
    \tau_{\mathrm{sfs},ij}^{\mathrm{mod,NL}} = \tau_{\mathrm{sfs},ij}^{\mathrm{G}} + \sigma^2 \left[ \dfrac{\partial \epsilon_{\mathrm{f}} \Bar{u}_i }{\partial x_k} + \sum_q v_{q,i} \dfrac{\partial \epsilon_{\mathrm{p},q}}{\partial x_k}  \right]\left[ \dfrac{\partial \epsilon_{\mathrm{f}} \Bar{u}_j }{\partial x_k} + \sum_q v_{q,j}\dfrac{\partial \epsilon_{\mathrm{p},q}}{\partial x_k}  \right].
\end{align}
We refer to this model as the non-linear model since it is a generalization of the model proposed by \citet{Leonard1975}. The non-linear model can be formulated as a tensorial particle viscosity model in analogy to a turbulent viscosity in turbulent single-phase flow. For comparison, we adapt the most common scalar turbulent viscosity model, the Smagorinsky model, to a volume-filtered particle-laden flow. The adaption is required to ensure Galilean invariance of the model. Using the identity equation \eqref{eq:switichingfilteringderivative}, we can write the modeled subfilter stress tensor analogously to the Smagorinsky model in turbulent single-phase flow
\begin{align}
\label{eq:smagorinsky}
    \tau_{\mathrm{sfs},ij}^{\mathrm{inter,Sm}} - \dfrac{1}{3}\tau_{\mathrm{sfs},kk}^{\mathrm{inter,Sm}}\delta_{ij} = - 2\nu_{\mathrm{p}}\Bar{S}_{ij},
\end{align}
with the additional viscosity that we refer to as particle viscosity
\begin{align}
    \nu_{\mathrm{p}} = (C_{\mathrm{s,p}}\sigma)^2\sqrt{2\Bar{S}_{ij}\Bar{S}_{ij}},
\end{align}
where $C_{\mathrm{s,p}}$ is a constant. The Galilean invariant volume-filtered strain-rate tensor is given by
\begin{align}
    \Bar{S}_{ij} = \dfrac{1}{2}\left[ \dfrac{\partial \epsilon_{\mathrm{f}} \Bar{u}_i }{\partial x_j} + \sum_q v_{q,i} \dfrac{\partial \epsilon_{\mathrm{p},q}}{\partial x_j} + \dfrac{\partial \epsilon_{\mathrm{f}} \Bar{u}_j }{\partial x_i} + \sum_q v_{q,j} \dfrac{\partial \epsilon_{\mathrm{p},q}}{\partial x_i}\right],
\end{align}
where the volume-filtered velocity gradients are replaced using equation \eqref{eq:switichingfilteringderivative}. \\
Note that it is equivalent to equation \eqref{eq:smagorinsky} if the strain-rate tensor of the volume-filtered velocity
\begin{align}
    \dfrac{1}{2}\left[ \dfrac{\partial \epsilon_{\mathrm{f}} \Bar{u}_i }{\partial x_j}  + \dfrac{\partial \epsilon_{\mathrm{f}} \Bar{u}_j }{\partial x_i} \right], \nonumber
\end{align}
is used together with the alternative viscous closure given in equation \eqref{eq:alternativeviscousclosure}.

\section{Evaluation of the closures}
\label{sec:evaluationclosures}

\subsection{Considered cases}
\begin{figure}
    \hspace{-0.1cm}
    \includegraphics[scale=0.45]{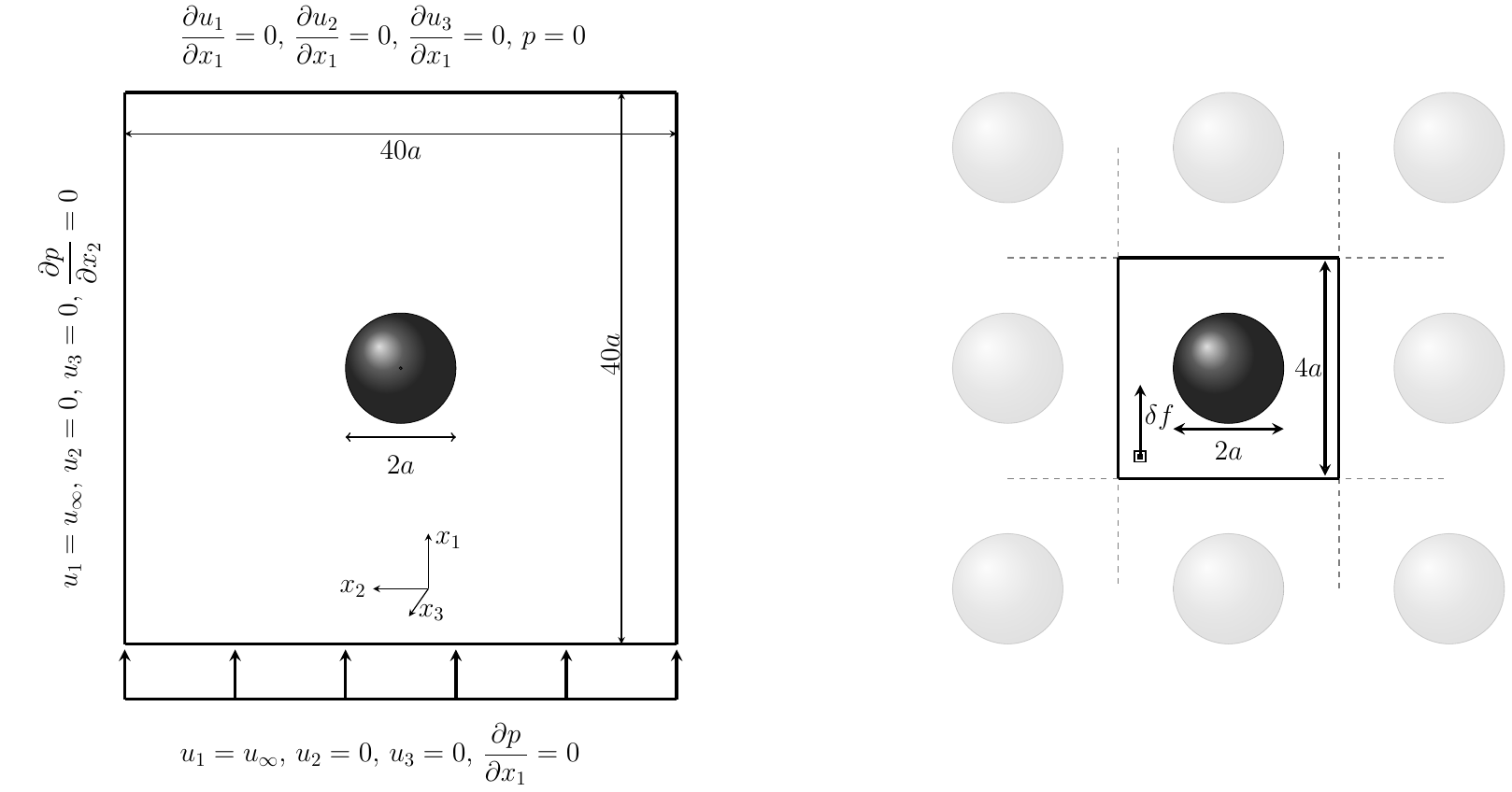}
    \caption{Sketch of the simulation domain of the single sphere configuration (left) and the periodic array of spheres (right).}
    \label{fig:domain}
\end{figure}

To test the implications of the previous section and to assess the importance of the different closures in the volume-filtered momentum equation, we study two different configurations of particles immersed in different flows. The first configuration is an isolated sphere in uniform flow at different Reynolds numbers. In Stokes flow, the existence of an analytical solution for this configuration allows us to study the effect of volume filtering in an infinite domain without a discretization error. We extend this configuration to higher Reynolds numbers Reynolds numbers ($\mathrm{Re}=20$ and $\mathrm{Re}=100$) and refer to this first configuration as single sphere configuration. With the second configuration, we study the influence of neighboring particles by simulating a sphere in a fully periodic computational domain, which we refer to as periodic array of spheres. \\
The cases of finite $\mathrm{Re}$ require numerical simulations, which are carried out with our in-house flow solver. The flow is solved with a finite-volume method, which employs momentum-weighted interpolation for a coupled solution of continuity and momentum that is second order in time and space \citep{Denner2020,Bartholomew2018}. The no-slip boundary condition at the particle surface is enforced with an immersed-boundary method (IBM) as described in \citet{AbdolAzis2019,Cheron2023a}.\citet{Cheron2023b} perform validations to obtain temporal and spatial resolutions that lead to converged results. In the present work we adapt these resolutions. \\
The left part of figure \ref{fig:domain} shows a sketch of the simulation domain of the single sphere configuration. The simulation domain is a cube of size $40a$ with the particle placed in the center and the flow is driven by a uniform inflow with the velocity $u_{\infty}$ at the lower face of the steamwise direction, $x_1$. The other velocities and the pressure gradient are zero. The same boundary conditions are applied at the faces in the two normal direction, $x_2$ and $x_3$. At the upper streamwise face, the velocity gradients and the pressure are set to zero. \\
The simulation domain of the periodic array of spheres is depicted in the right part of figure \ref{fig:domain}. The cubic periodic simulation domain has a size of $4a$, which corresponds to a global particle volume fraction of $\epsilon_{\mathrm{p,glob}}=0.065$. The flow is driven by a constant volume force, $\delta f$, applied in the streamwise direction. The Reynolds number of the periodic array of spheres is defined as
\begin{align}
    \mathrm{Re}_{\mathrm{periodic}} = \dfrac{\rho_{\mathrm{f}}|\boldsymbol{u}_{\mathrm{sf}}|2a}{\mu_{\mathrm{f}}},
\end{align}
where the superficial velocity is given by
\begin{align}
    u_{\mathrm{sf},i} = \dfrac{1}{V_{\mathrm{periodic}}}\int \displaylimits_{\Omega_{\mathrm{periodic}}} I_{\mathrm{f}}u_i \mathrm{d}V.
\end{align}
The volume of the periodic simulation domain $\Omega_{\mathrm{periodic}}$, which includes the volume occupied by the fluid and the particle, is indicated with $V_{\mathrm{periodic}}$.

\subsection{Stokes flow around a single sphere}
\label{sec:Stokesflowsinglesphere}
\begin{figure}
    \centering
    \hspace{-1.7cm}
    \includegraphics[scale=0.75]{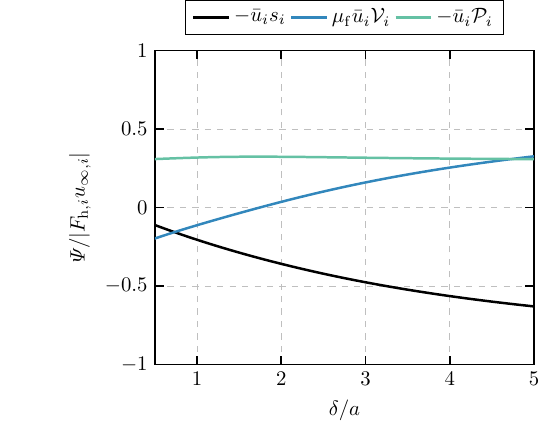}
    \caption{Volume integrated energy transfer rates of the flow over a fixed sphere in Stokes flow. The energy transfer rates of the particle momentum source, the viscous term, and the pressure term are plotted.}
    \label{fig:energytransferstokesv0}
\end{figure}

\begin{figure}
    \centering
    \hspace{-1.7cm}
    \includegraphics[scale=0.75]{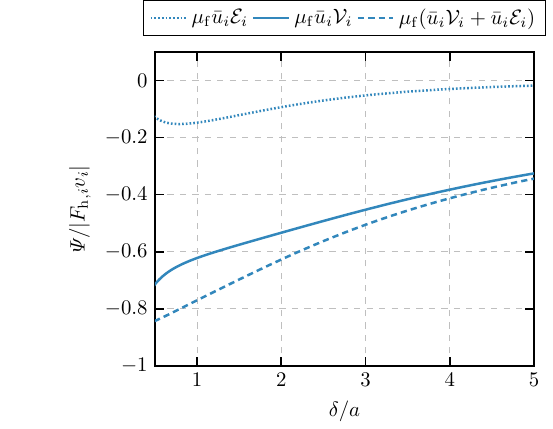}
    \caption{Volume integrated energy transfer rates of a sphere moving in quiescent fluid in the Stokes regime. The energy transfer rates of the viscous closure term, the viscous term, and their sum are plotted.}
    \label{fig:energytransferstokesv-1}
\end{figure}

\begin{figure}
    \centering
    \hspace{0.4cm}
    \includegraphics[scale=0.6]{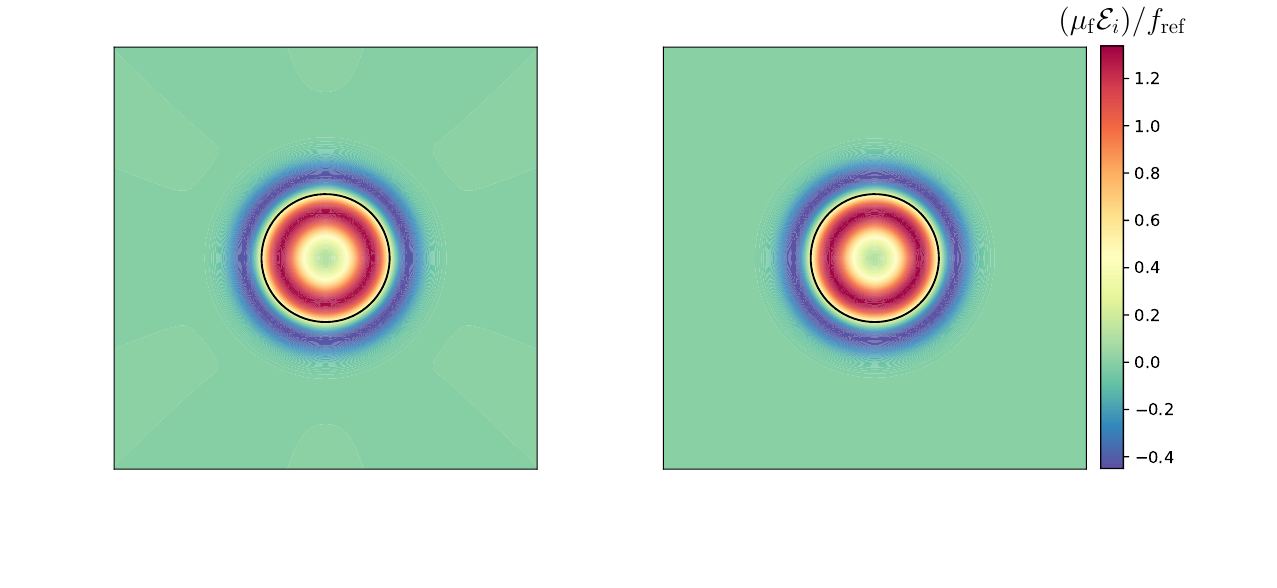}
    \caption{Contours of the viscous closure at filter width $\delta/a= 1$. Left shows the closure by explicitly subtracting the stresses and right shows the closure that is derived analytically in section \ref{ssec:viscousclosure}. The black circle indicates the surface of the particle.}
    \label{fig:contourviscousclosure}
\end{figure}

\begin{figure}
    \centering
    \hspace{-1.3cm}
    \includegraphics[scale=0.75]{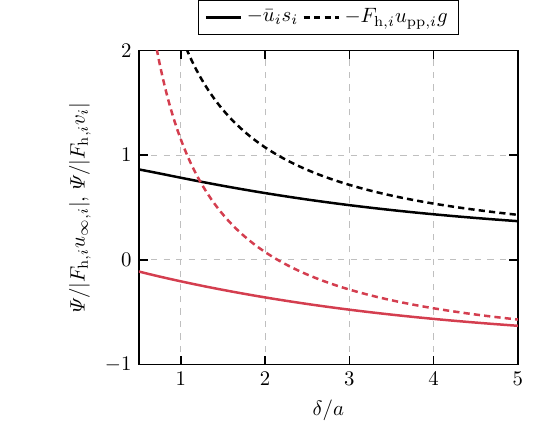}
    \caption{Volume integrated energy transfer rates of the analytical particle momentum source and the particle momentum source in the simplified point-particle framework in Stokes flow. Black indicates configuration of the moving sphere in a quiescent fluid and red indicates the flow over a fixed sphere.}
    \label{fig:regularizedsourceoverfilterwidth}
\end{figure}

We consider the Stokes flow around a sphere in an infinite domain. In one configuration the particle steadily moves, with a constant velocity, in the negative streamwise direction in a quiescent fluid and in the other configuration the particle is fixed but there is a constant positive fluid velocity. Although these cases are analogous, by considering a different reference frame in which the sphere moves, the viscous closure, $\mathcal{E}$, may be investigated, which vanishes if the sphere does not move with respect to the reference frame.\\
We can define a kinetic energy of the volume-filtered fluid velocity $K_{\mathrm{VF}}=\epsilon_{\mathrm{f}}\Bar{u}_i\Bar{u}_i/2$. By taking the dot product of the volume-filtered fluid momentum equation \eqref{eq:momentumwithclosures} with $\Bar{u}_i$, we obtain a transport equation for the kinetic energy of the volume-filtered fluid velocity
\begin{align}
\label{eq:volumefilteredkineticenergyequation}
    \rho_{\mathrm{f}}\dfrac{\partial K_{\mathrm{VF}}}{\partial t} + \Bar{u}_i\rho_{\mathrm{f}}\dfrac{\partial}{\partial x_j}(\epsilon_{\mathrm{f}}\Bar{u}_i\epsilon_{\mathrm{f}}\Bar{u}_j) = - \Bar{u}_i\dfrac{\partial \epsilon_{\mathrm{f}}\Bar{p}}{\partial x_i}+\mu_{\mathrm{f}}  \Bar{u}_i\dfrac{\partial^2\epsilon_{\mathrm{f}}\Bar{u}_i}{\partial x_j \partial x_j} -  \Bar{u}_i s_i+ \nonumber \\
    \mu_{\mathrm{f}}  \Bar{u}_i\mathcal{E}_i - \rho_{\mathrm{f}} \Bar{u}_i\dfrac{\partial }{\partial x_j}\tau_{\mathrm{sfs},ij}.
\end{align}
In the limit of steady Stokes flow the transport equation reduces to 
\begin{align}
    0 = - \Bar{u}_i\dfrac{\partial \epsilon_{\mathrm{f}}\Bar{p}}{\partial x_i}+\mu_{\mathrm{f}}  \Bar{u}_i\dfrac{\partial^2\epsilon_{\mathrm{f}}\Bar{u}_i}{\partial x_j \partial x_j} -  \Bar{u}_i F_{\mathrm{h},i}\mathcal{K}_{\mathrm{St}}(|\boldsymbol{x}-\boldsymbol{x}_{\mathrm{p}}|)+ \mu_{\mathrm{f}}  \Bar{u}_i\mathcal{E}_i.
\end{align}
For conciseness we abbreviate the pressure term $\mathcal{P}_i=\dfrac{\partial \epsilon_{\mathrm{f}}\Bar{p}}{\partial x_i}$ and the viscous term $\mathcal{V}_i=\dfrac{\partial^2\epsilon_{\mathrm{f}}\Bar{u}_i}{\partial x_j \partial x_j}$. In figure \ref{fig:energytransferstokesv0} the different volume integrated energy transfer rates $\varPsi$ that occur in the kinetic energy transport equation are plotted for the flow around a fixed sphere. The energy transfer rate of a quantity $\varPhi_i$ in the volume-filtered fluid momentum equation \eqref{eq:momentumwithclosures} is given as 
\begin{align}
    \varPsi = \int \displaylimits_{\Omega_{\mathrm{int}}}\Bar{u}_i \varPhi_i \mathrm{d}V,
\end{align}
where $\Omega_{\mathrm{int}}$ is the integration region, which corresponds to a cube with an edge length of ten particle diameters, i.e., $20a$. The trends with respect to the filter width of the presented results are verified to be insensitive to $\Omega_{\mathrm{int}}$, although the magnitudes of some energy transfer rates change. Therefore, the results must be interpreted as the energetic effect of particles in an influence region $\Omega_{\mathrm{int}}$, which is the most interesting region because the energy transfer rates possess the largest magnitudes there. \\
When the particle is fixed and the filter width $\delta=0$, no kinetic energy is exchanged between fluid and particle because the velocity at the particle surface is zero. The energy transfer rate of the particle momentum source $\varPsi=-F_{\mathrm{h},i}v_i$ is zero. With increased filter width, the energy transfer rate is negative. Consequently, the  energy transfer rate of the particle momentum source removes kinetic energy from the filtered scales and adds kinetic energy to the subfilter scales, such that the total energy transfer rate of all scales equals zero. This is in alignment with the observation that particles in turbulence remove kinetic energy from large scales and add them to smaller scales \citep{Squires1994,Sundaram1999,Ferrante2003,Mallouppas2017}. \\
We further observe that the energy transfer rate induced by the pressure term remains relatively constant as the filter width increases indicating that the pressure energy exchange takes place at large scales. The energy transfer rate of the viscous term possesses its minimum at $\delta=0$. Most of the viscous dissipation takes place at small filter width. For larger filter widths the viscous source term even becomes positive, which seems counter intuitive. The viscous energy source term can be decomposed as
\begin{align}
    \int\displaylimits_{\Omega}\mu_{\mathrm{f}}  \Bar{u}_i\dfrac{\partial^2\epsilon_{\mathrm{f}}\Bar{u}_i}{\partial x_j \partial x_j}\mathrm{d}V = \int\displaylimits_{\Omega}\dfrac{\mu_{\mathrm{f}}}{2\epsilon_{\mathrm{f}}} \dfrac{\partial^2 (\epsilon_{\mathrm{f}}\Bar{u}_i)^2}{\partial x_j \partial x_j}\mathrm{d}V - \int\displaylimits_{\Omega}\dfrac{\mu_{\mathrm{f}}}{\epsilon_{\mathrm{f}}}\dfrac{\partial \epsilon_{\mathrm{f}}\Bar{u}_i}{\partial x_j}\dfrac{\partial \epsilon_{\mathrm{f}}\Bar{u}_i}{\partial x_j}\mathrm{d}V.
\end{align}
The second term on the right-hand side is always negative and analogous to the viscous fluid flow dissipation which is also present in a single-phase flow. In the fixed particle configuration, the curvature of the squared volume-filtered velocity is positive and the local fluid volume fraction is smaller than one. Consequently, positive curvatures of the squared volume-filtered velocity receive a larger weight than negative curvatures, and the first term on the right-hand side is positive and even outweighs the second term for large filter widths. \\
In figure \ref{fig:energytransferstokesv-1}, the frame of reference is shifted, such that the particle moves in a quiescent fluid. The energy transfer rate induced by the viscous closure is compared to the energy transfer rate of the total viscous term. The largest contribution of the viscous closure to the energy transfer rate is at $\delta \approx a$, where it makes up approximately 20\% of the energy transfer rate of the viscous term. For larger filter widths the energy transfer rate induced by the viscous closure approaches zero. In the fixed sphere configuration the closure is zero because the particle translational velocity is zero, i.e., the volume fraction does not change. \\
We consider the moving particle in the Stokes flow to evaluate the spatial distribution of the viscous closure $\mathcal{E}_i$. Figure \ref{fig:contourviscousclosure} compares the contours in a plane normal to the streamwise direction of the viscous closure computed by subtracting the viscous stresses as in equation \eqref{eq:definitionviscousclosure} using quantities not accessible in the volume-filtered framework and computed using the exact relation of equation \eqref{eq:viscousclosure} containing only quantities that are known in the volume-filtered framework. The results are normalized with $f_{\mathrm{ref}}=|\boldsymbol{F}_{\mathrm{h}}|/V_{\mathrm{p}}$. The agreement of the two contours confirms the derivation in section \ref{ssec:viscousclosure}. Note that $\mathcal{E}_i$ is spherically symmetric, which is always guaranteed for spherical particles. Since the particle volume fraction corresponding to a spherical particle is spherical symmetric, its Laplacian possesses spherical symmetry too. The derivation of the expression for $\mathcal{E}_i$ shows that it is valid also outside the Stokes regime and in the vicinity of other particles. \\
A very common simplification of the volume-filtered Euler-Lagrange framework is to use the following fluid momentum equation instead of equation \eqref{eq:momentumwithclosures} \citep{Maxey2017,Maxey2001,Evrard2020a,Poustis2019}:
\begin{align}
\label{eq:simplifiedEL}
    \rho_{\mathrm{f}}\dfrac{\partial u_{\mathrm{pp},i}}{\partial t} + \rho_{\mathrm{f}}\dfrac{\partial  u_{\mathrm{pp},i}u_{\mathrm{pp},j}}{\partial x_j} = -\dfrac{\partial p_{\mathrm{pp}}}{\partial x_i} + \mu_{\mathrm{f}} \dfrac{\partial^2 u_{\mathrm{pp},i}}{\partial x_j \partial x_j} - F_{\mathrm{h},i}g(|\boldsymbol{x}-\boldsymbol{x}_p|),
\end{align}
where the index $\mathrm{pp}$ indicates a quantity in the framework that we refer to as the simplified point-particle framework, and $g$ is a Gaussian regularization kernel representing the fluid-particle momentum exchange. This simplification is typically used for dilute regimes. We neglect the assumption of a constant fluid volume fraction $\epsilon_{\mathrm{f}}$ and focus on the last term on the right-hand side of equation \eqref{eq:simplifiedEL}, the particle momentum exchange. We simplify our discussion by considering the Stokes flow around an isolated spherical particle
\begin{align}
    0 = -\dfrac{\partial p_{\mathrm{pp}}}{\partial x_i} + \mu_{\mathrm{f}} \dfrac{\partial^2 u_{\mathrm{pp},i}}{\partial x_j \partial x_j} - F_{\mathrm{h},i}g(|\boldsymbol{x}-\boldsymbol{x}_p|).
\end{align}
Since the volume integrals of the Gaussian $g$ and the analytical kernel in the Stokes limit $\mathcal{K}_{\mathrm{St}}$ equal one, the total momentum together with the particle is conserved.
However, it has been argued in previous studies that the simplified momentum equation \eqref{ssec:viscousclosure} is not energetically consistent if $\int u_{\mathrm{pp},i}F_{\mathrm{h},i}g(|\boldsymbol{x}-\boldsymbol{x}_p|)\mathrm{d}V\not= v_i F_{\mathrm{h},i}$ \citep{Xu2007a,Evrard2020a,Frohlich2018}, where energetical consistency is satisfied if viscous dissipation or inelastic collisions are the only mechanisms to reduce the total kinetic energy of the fluid-particle mixture. The simplified framework is indeed not energy consistent, but for a different reason. The reason for the inconsistency is that $\int u_{\mathrm{pp},i}F_{\mathrm{h},i}g(|\boldsymbol{x}-\boldsymbol{x}_p|)\mathrm{d}V\not=\int \Bar{u}_is_i\mathrm{d}V$, i.e., a Gaussian is, strictly speaking, not the right kernel to regularize the particle momentum source. \\
Figure \ref{fig:regularizedsourceoverfilterwidth} shows the energy transfer rate of the particle momentum exchange term for the filtered momentum equation and the simplified momentum equation in Stokes flow. The results are shown for the moving sphere and the fixed sphere in the Stokes regime. The results of the simplified momentum equation are obtained analytically using the solution of a regularized Stokeslet (see, e.g., \citet{Cortez2001}). \\
As the filter width approaches zero, the explicitly volume-filtered energy transfer terms converge towards $ v_i F_{\mathrm{h},i}$. In the simplified point-particle framework, the energy transfer term becomes infinitely large. \citet{Evrard2020a} show that, for a filter width of $\delta/a\approx 2{.}148$, the energy transfer equals the energy that is transferred to (or removed from) the particles $ v_i F_{\mathrm{h},i}$ in the Stokes regime, if a Wendland kernel is employed to regularize the momentum source, which is a polynomial approximation of the Gaussian kernel. Contrary to what is claimed in \citet{Evrard2020a}, $\delta/a\approx 2{.}148$ does not correspond to a consistent energy transfer because this would assume that the complete energy exchange with the particle takes only place at the filtered scales. In fact, the energy transfer is consistent if, and only if, the energy transfer term equals the the energy transfer rate from the explicit filtering. However, it is observed in figure \ref{fig:regularizedsourceoverfilterwidth}, that the energy transfer of the simplified point-particle framework approaches the explicitly filtered solution with increasing filter width. This is supported by the finding that $\mathcal{K}_{\mathrm{St}}\approx g$ for $\delta\gg1$. \\
It is worth highlighting that instabilities in the simplified point-particle framework can occur for small filter widths, since the Gaussian approaches a singular Dirac delta distribution as it is observed in \citet{Cheron2023}.


\subsection{Single sphere at higher Reynolds number}
We split the analysis of the single sphere in a flow with finite Reynolds number into an a priori analysis where the explicitly volume-filtered particle-resolved DNS, i.e., a simulation that resolves the small flow structures around the particle, is investigated and an a posteriori analysis where the volume-filtered Navier-Stokes equations are solved with the proposed models for the closures and compared to the explicitly volume-filtered particle-resolved DNS. 

\subsubsection{A priori analysis of the energy transfer}
\label{sec:apriorienergytransfer}
\begin{figure}
    \centering
    \hspace{-1.7cm}
    \includegraphics[scale=0.75]{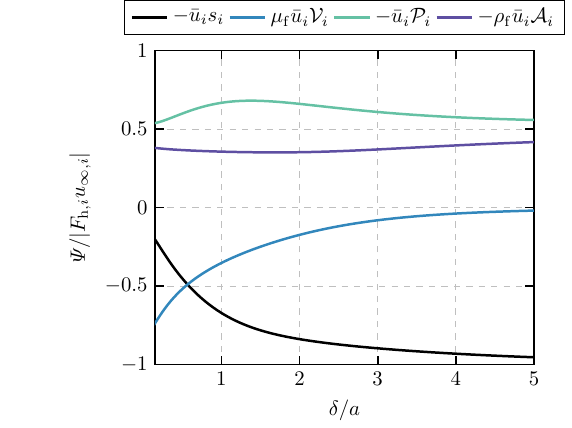}
    \caption{Volume integrated energy transfer rates of the flow over a fixed sphere at $Re=100$. The energy transfer rates of the particle momentum source, the viscous term, the pressure term, and the advective term are plotted.}
    \label{fig:energytransferRe100v0}
\end{figure}

Similar to the studies in Stokes flow outlined earlier, we consider the flow around a single sphere at $\mathrm{Re}=100$ in two frames of reference, a fixed sphere in a uniform flow and a moving sphere in a quiescent fluid. The purpose of the moving sphere configuration is to investigate the influence of the viscous closure, $\mathcal{E}_i$, at higher $\mathrm{Re}$, which is zero for a fixed particle.  \\
Figure \ref{fig:energytransferRe100v0} shows the volume integrated terms of the kinetic energy equation \eqref{eq:volumefilteredkineticenergyequation} for the fixed sphere configuration. The volume for the integration is a cube with an edge length of ten particle diameters. Note that $\mathcal{A}_i=\dfrac{\partial \epsilon_{\mathrm{f}}\overline{u_i u_j}}{\partial x_j}$ is an abbreviation for the explicitly volume-filtered advective term. Similar to the Stokes flow around an isolated spherical particle, the energy transfer rate of the particle momentum source is zero for $\delta=0$ and becomes increasingly negative with increasing filter width. The particle momentum source removes energy from the filtered scales and adds energy to the subfilter scales. For the larger $\mathrm{Re}$ than in the Stokes limit, however, the energy transfer takes place at much smaller filter widths. This can be concluded because the slope of the energy transfer rate of the particle momentum source in figure \ref{fig:energytransferRe100v0} is very steep for small filter widths and flat for larger filter widths compared to the Stokes flow around the sphere. The energy transfer rate of the viscous term possesses the minimum at $\delta=0$ and approaches zero as the filter width increases. Consequently, the viscous dissipation predominates at the small scales. The energy transfer rate of the pressure term is positive for all filter widths. That means that at all filter widths, the pressure term adds energy to the filtered scales. The same is observed for the energy transfer rate of the advective term. However, it is observed that both energy transfer rates are not monotonically increasing or decreasing. The pressure term possesses a maximum and the advective term a minimum of energy transfer rate in the region of $1< \delta/a < 2$. 
Filtering out the small scale contribution of the energy transfer rate of the pressure term or the advective term in regions with positive slope of the pressure term or the advective term in figure \ref{fig:energytransferRe100v0} increases the respective energy transfer rate with respect to a smaller filter width. A negative slope of the pressure term or the advective term decreases the energy transfer rate with respect to a smaller filter width. Therefore, the pressure term and the advective term remove energy in regions of positive slope and add energy in regions of negative slop. The advective term removes energy from the large scales and adds energy to the small scales, which is qualitatively similar to what is known for single-phase turbulence (see, e.g., \citet{Pope2000}). The results for $\mathrm{Re}=20$ are qualitatively similar and, therefore, not shown. \\

\subsubsection{A priori analysis of the viscous closure}
\label{sec:aprioriviscousclosure}
\begin{figure}
    \centering
    \hspace{-1.7cm}
    \includegraphics[scale=0.75]{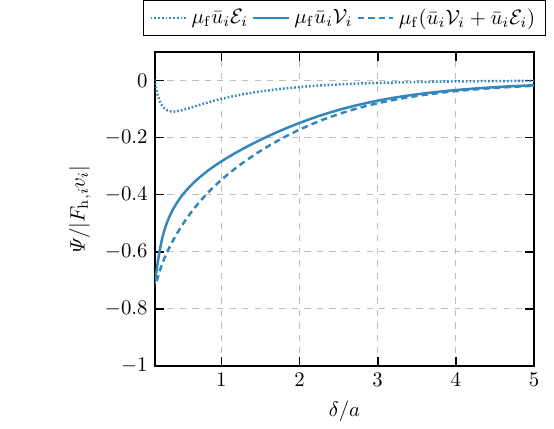}
    \caption{Volume integrated energy transfer rates of a sphere moving in quiescent fluid at $Re=100$. The energy transfer rates of the viscous closure term, the viscous term, and their sum are plotted.}
    \label{fig:energytransferRe100v-1}
\end{figure}

In figure \ref{fig:energytransferRe100v-1}, the energy transfer rate of the viscous closure, $\mu_{\mathrm{f}}\Bar{u}_i\mathcal{E}_i$, is shown for the case of $\mathrm{Re}=100$. Note that the frame of reference is shifted such that the particle moves with the velocity of $v_i$ in a quiescent fluid to obtain a non-zero viscous closure. Along with the energy transfer rate of the viscous closure, in figure \ref{fig:energytransferRe100v-1} we show the energy transfer rate of the viscous term, as it occurs in equation \eqref{eq:volumefilteredkineticenergyequation}, and the sum of both terms, which corresponds to the energy transfer rate of the explicitly volume filtered viscous term in the NSE. In alignment with the observations in Stokes flow, the energy transfer rate of the viscous closure reaches a maximum at filter widths $\delta<1$ and approaches zero as the filter width increases. Consequently, the energy transfer rate of the viscous term deviates from the energy transfer rate of the explicitly filtered viscous term at small filter widths, but both approach each other for increasing filter widths. Although the viscous closure has negligible impact on the energy transfer for a large filter width in the investigated cases, common configurations exist where the viscous closure has a significant influence, even for large filter widths. In a channel flow, for instance, the fluid and the particles move with a large velocity relative to the frame of reference. The large particle velocity in the frame of reference increases the magnitude of the viscous closure compared to the present configuration. Since we provide an analytical expression for the viscous closure, its closure is recommended especially in such configurations. \\

\subsubsection{A priori analysis of the subfilter stresses}
\label{sec:apriorisubfilterstresses}
\begin{figure}
    \hspace{-1.0cm}
    \includegraphics[scale=0.75]{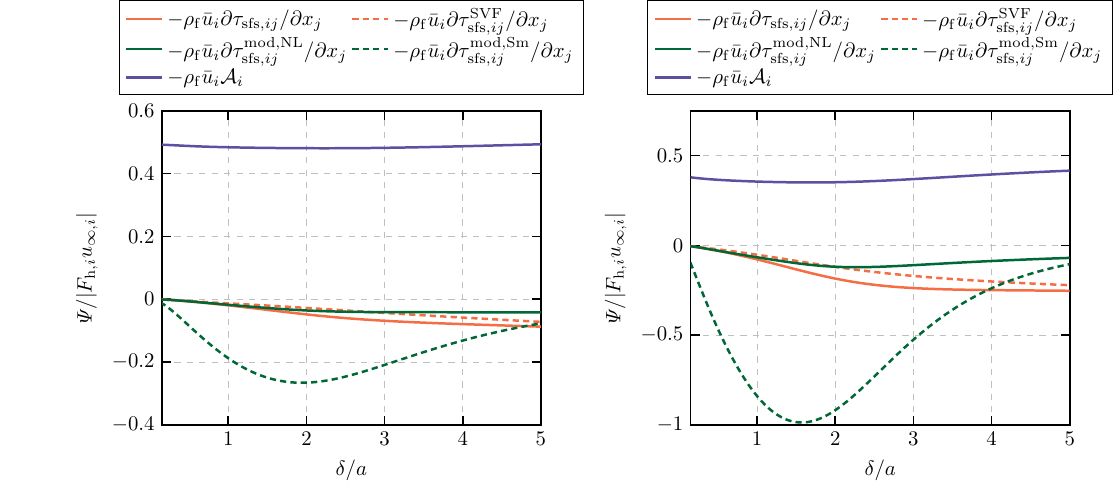}
    \caption{Volume integrated energy transfer rates of the flow over a fixed sphere at $\mathrm{Re}=20$ (left) and $\mathrm{Re}=100$ (right).  The energy transfer rates of the advective term, $\rho_{\mathrm{f}}\Bar{u}_i \mathcal{A}_i$, the subfilter stress tensor, $\rho_{\mathrm{f}}\Bar{u}_i \partial \tau_{\mathrm{sfs},ij}/\partial x_j$, the subfilter stress tensor with the single volume fraction definition, $\rho_{\mathrm{f}}\Bar{u}_i \partial \tau_{\mathrm{sfs},ij}^{\mathrm{SVF}}/\partial x_j$, the modeled subfilter stress tensor with the non-linear model, $\rho_{\mathrm{f}}\Bar{u}_i \partial \tau_{\mathrm{sfs},ij}^{\mathrm{mod,NL}}/\partial x_j$, and the subfilter stress tensor with the adapted Smagorinsky model, $\rho_{\mathrm{f}}\Bar{u}_i \partial \tau_{\mathrm{sfs},ij}^{\mathrm{mod,Sm}}/\partial x_j$, are plotted.}
    \label{fig:advectivweenergytransfer}
\end{figure}

\begin{figure}
    \hspace{1cm}
    \includegraphics[scale=0.6]{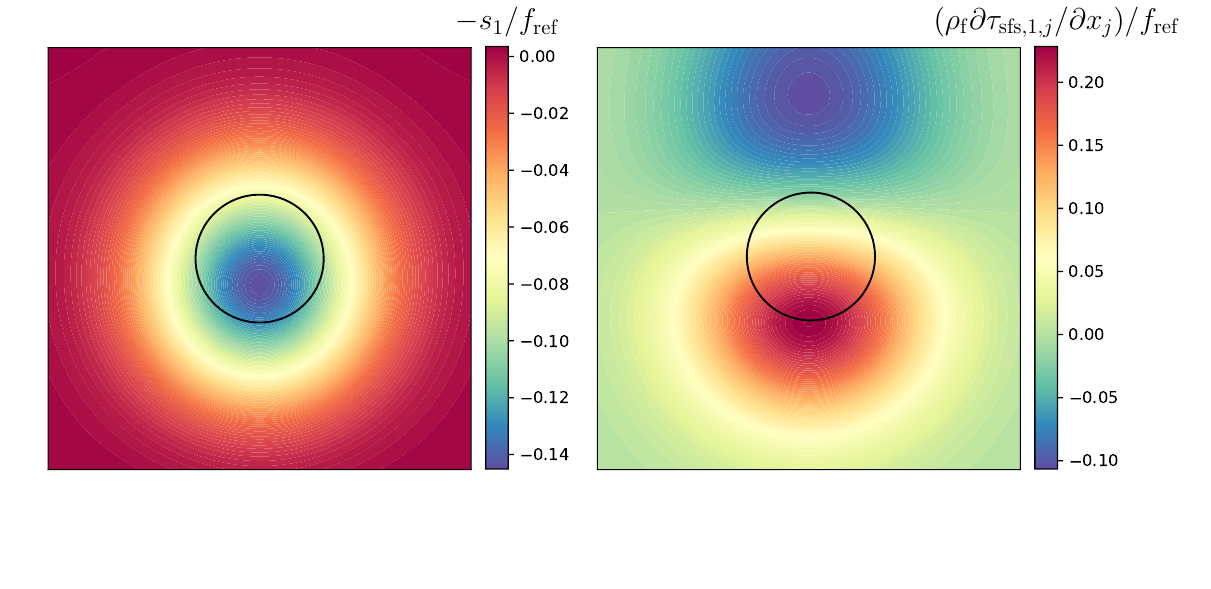}
    \caption{Contours of the momentum contributions in the streamwise direction of the particle momentum source (left) and the divergence of the subfilter stress tensor (right) for the fixed sphere at $\mathrm{Re}=100$ and filter width $\delta/a=4$.}
    \label{fig:momentumtauRe100v0d4}
\end{figure}

For the study of the subfilter stresses, we consider the frame of reference where the particle is fixed, since the subfilter stress tensor is non-zero also in this configuration. In figure \ref{fig:advectivweenergytransfer} the contributions of the subfilter stress tensor to the energy transfer are shown. In addition to the subfilter stress tensor, we also plot the energy transfer rate of the non-linear model that is proposed in section \ref{sec:modeltau}, the Smagorinsky model adapted to a particle-laden flow, and the subfilter stress tensor as it would be defined if the single volume fraction advective term $\tau_{\mathrm{sfs},ij}^{\mathrm{SVF}}$ would be used. Furthermore, the energy transfer rate of the explicitly filtered advective term is shown for comparison. In the left part of figure \ref{fig:advectivweenergytransfer}, the results of $\mathrm{Re}=20$ are shown and in the right part the results of $\mathrm{Re}=100$ are shown. For $\delta=0$ and both $\mathrm{Re}$, the energy transfer rate of the subfilter stress tensor and the subfilter stress tensor with the single volume fraction definition of the advective term are zero. With increasing filter width, both terms yield increasingly negative energy transfer rates. With both definitions, $\tau_{\mathrm{sfs},ij}$ and  $\tau_{\mathrm{sfs},ij}^{\mathrm{SVF}}$, the subfilter stress tensor removes energy from the filtered scales and adds energy to the subfilter scales. This is analog to the behavior of the subfilter stress tensor in LES of single-phase turbulence, where the unresolved turbulent subfilter scales have a mainly dissipative effect on the filtered scales. This motivates the use of an additional (typically positive) turbulent viscosity as model for the subfilter stress tensor. The energy transfer rates of both subfilter stress tensor definitions and both investigated $\mathrm{Re}$ do not approach zero as the filter width increases. At the largest investigated filter width, $\delta/a=5$, the energy transfer rates of the subfilter stress tensors make up approximately 50\% of the energy transfer rate of the advective term for $\mathrm{Re}=100$. For $\mathrm{Re}=20$, however, both definitions of the subfilter stress tensors contribute only with approximately 20\% to the energy transfer rate of the advective term. The significant contribution of the subfilter stress tensor to the energy transfer suggest that it requires modeling for the investigated configuration of an isolated sphere, even at larger filter widths. Note that even at a small energy transfer rate of the subfilter stress tensor, its contribution to the momentum can be significant. Figure \ref{fig:momentumtauRe100v0d4} compares the contributions to the fluid momentum equation of the particle momentum source and the subfilter stress term in the streamwise direction for the flow over a fixed sphere at $\mathrm{Re}=100$ and filter width $\delta/a=4$. The magnitude of the subfilter stress term is of similar magnitude and even exceeds the magnitude of the particle momentum source, which clearly highlights that the subfilter stress tensor requires modeling. In this particular configuration, neglecting the subfilter stress tensor would be as severe as omitting the particle momentum source. \\
\begin{figure} 
    \centering
    \hspace{0.4cm}
    \includegraphics[scale=0.6]{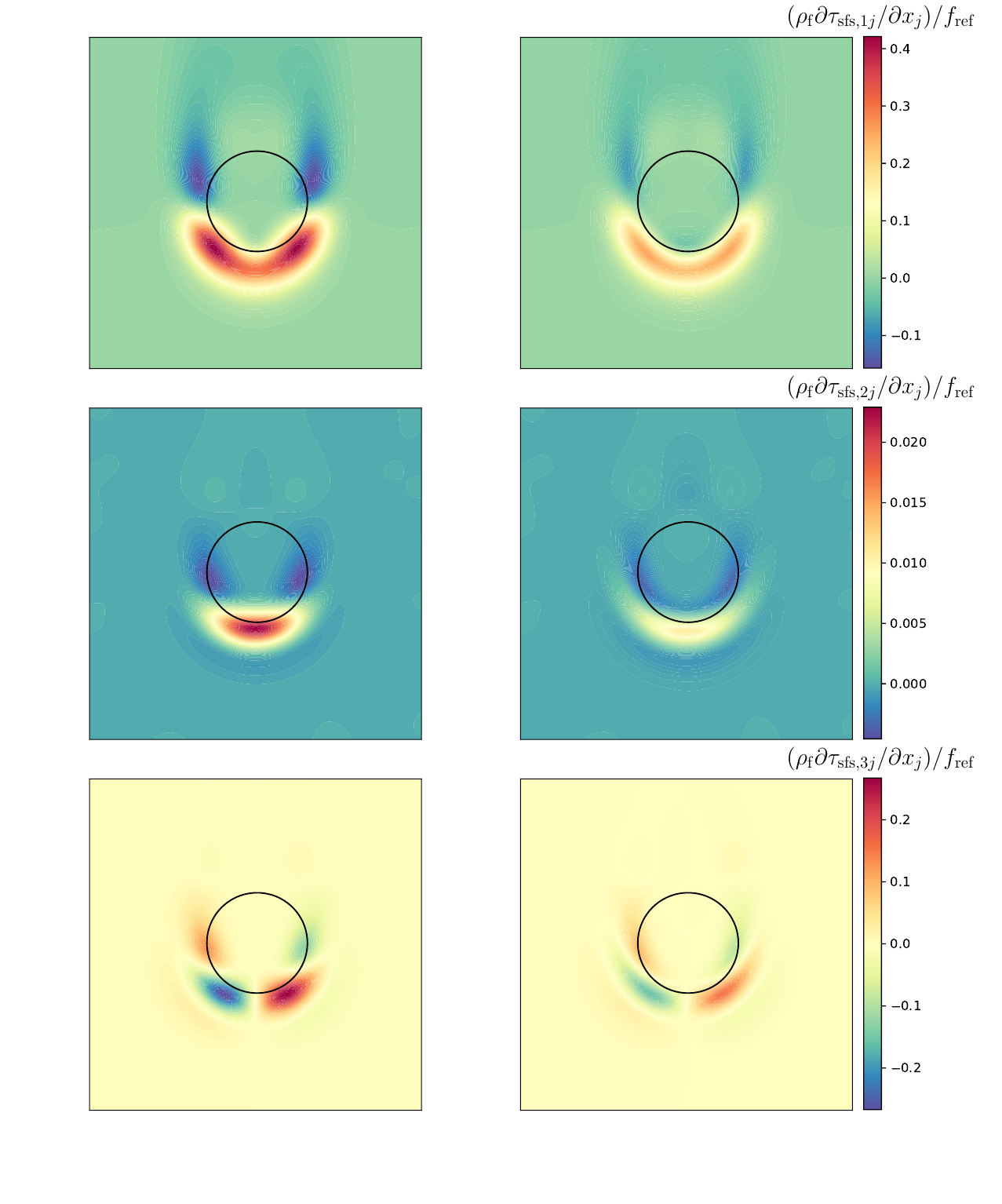}
    \caption{Divergence of the subfilter stress tensor of a flow around a fixed sphere at $\mathrm{Re}=100$ and for the filter width. The plots on the left are obtained from explicitly volume-filtering the the particle-resolved simulation and the plots on the right are computed with the non-linear model. The filter width of the shown case is $\delta/a=1$}
    \label{fig:subfilterstressclosure}
\end{figure}
We also show the energy transfer rate of the non-linear model for the subfilter stress tensor, that we propose in section \ref{sec:modeltau}, and the Smagorinsky model adapted to a particle-laden flow in figure \ref{fig:advectivweenergytransfer}. Up to a filter width of $\delta/a\approx1$, the energy transfer rate of the non-linear model almost perfectly agrees with the energy transfer rate of the subfilter stress tensor for $\mathrm{Re}=20$ and $\mathrm{Re}=100$. As the filter width increases further, however, the agreement deteriorates. The same behavior is observed in single-phase turbulence \citep{Borue1998,Johnson2021}. For larger filter widths, the neglected integral terms in equation \eqref{eq:formalsolutiontau} predominate. The Smagorinsky model yields an energy transfer rate of a too large magnitude for $\delta/a<4$. At larger filter widths, the magnitude of the energy transfer rate is too small. Although the Smagorinsky model contains a constant, that we assumed to be $C_{\mathrm{s,p}}=1$ in the present case, no constant exists that leads to a matching energy transfer rate for a wide range of filter widths, which is clearly observed in figure \ref{fig:advectivweenergytransfer} as the energy transfer rate by the Smagorinsky model behaves differently with respect to the filter width than the energy transfer rate of the subfilter stress tensor. Therefore, the Smagorinsky model can generally not be recommended as a model for the subfilter stress tensor; at least not for the flow around an isolated sphere. \\
Figure \ref{fig:subfilterstressclosure} shows contours of the three components of the divergence of the subfilter stress tensor for the subfilter stress tensor and the non-linear model proposed in section \ref{sec:modeltau} for a filter width of $\delta/a=1$ and $\mathrm{Re}=100$. The results are normalized with $f_{\mathrm{ref}}=|\boldsymbol{F}_{\mathrm{h}}|/V_{\mathrm{p}}$. The shape of the contours of the modeled subfilter stress tensor show similar structures as the contours of the subfilter stress tensor. The first component, which is the streamwise component, possesses the largest magnitude. The contours are very similar in shape but the magnitudes deviate. The second component possesses the smallest magnitude. Although the overall shape of the contours is similar, minor deviations in shape occur near the upstream surface of the sphere. The third component agrees well in shape but shows the same discrepancies in magnitude as the other components. For smaller filter widths, the agreement in shape and magnitude improves, but for larger filter widths the modeled subfilter stress tensor shows larger deviations, which supports the observations of the energy transfer rates.

\subsubsection{A priori analysis of the particle momentum source}
\label{sec:aprioriparticlemomentumsource}
\begin{figure}
    \centering
    \includegraphics[scale=0.75]{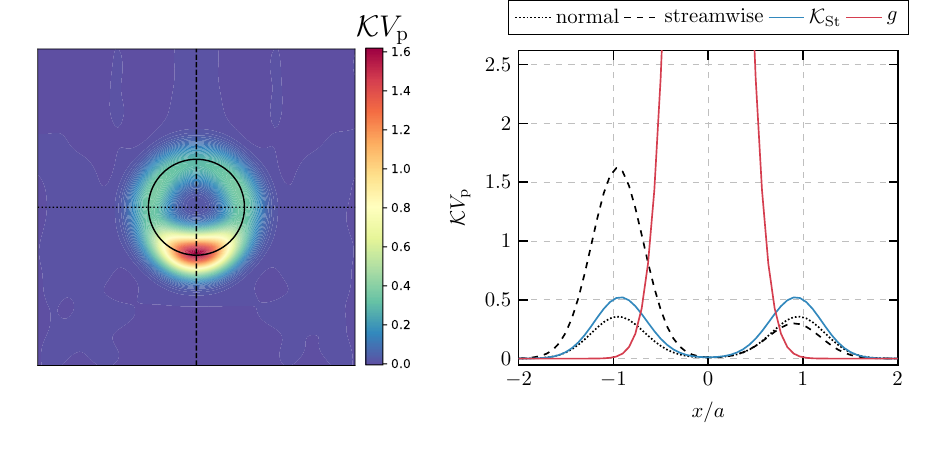}
    \caption{Shape of the particle momentum source for a flow over a fixed sphere at $\mathrm{Re}=100$ and filter width $\delta/a=1$. The left plot shows a contour of the regularization kernel. The right plot shows the regularization kernel in the streamwise and normal direction together with the analytical regularization kernel for Stokes flow and a Gaussian of the same filter width.}
    \label{fig:momentumsourceshape1Re100}
\end{figure}

\begin{figure}
    \centering
    \includegraphics[scale=0.75]{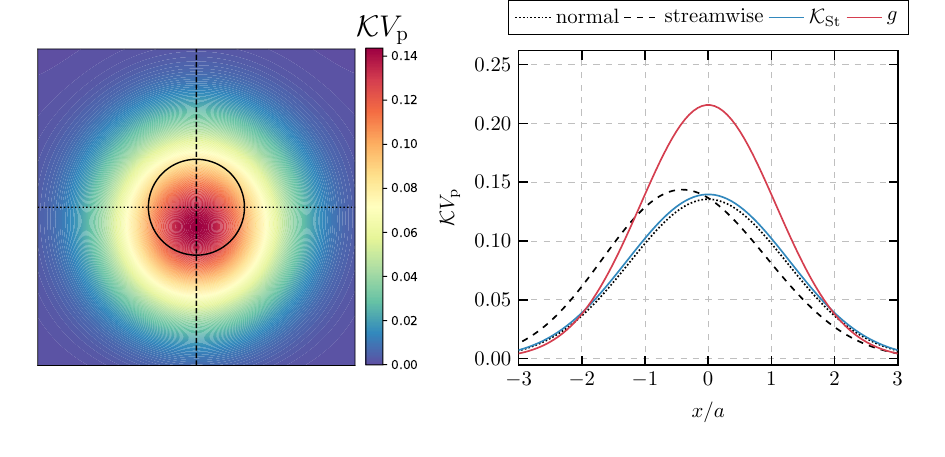}
    \caption{Shape of the particle momentum source for a flow over a fixed sphere at $\mathrm{Re}=100$ and filter width $\delta/a=4$. The left plot shows a contour of the regularization kernel. The right plot shows the regularization kernel in the streamwise and normal direction together with the analytical regularization kernel for Stokes flow and a Gaussian of the same filter width.}
    \label{fig:momentumsourceshape4Re100}
\end{figure}

In figure \ref{fig:momentumsourceshape1Re100} the spatial distribution of the particle momentum source in a flow of $\mathrm{Re}=100$ is compared to a Gaussian and the analytical kernel in Stokes flow $\mathcal{K}_{\mathrm{St}}$ for a filter width $\delta/a=1$. The spatial distribution of the particle momentum source appears as a ring in the two-dimensional slice, and is a spherical hull in three dimensions. In contrast to the spherical symmetry of the kernel in the Stokes limit, an upstream-downstream asymmetry is observed for larger $\mathrm{Re}$. Near the stagnation point, the greatest magnitude of the kernel is observed. In the right part of figure \ref{fig:momentumsourceshape1Re100}, the kernel is plotted along lines through the center of the sphere in the streamwise and normal direction together with the analytical kernel in the Stokes limit and the commonly used Gaussian regularization of the same filter width. The magnitude of the kernel near the stagnation point is approximately six times as large as the magnitude on the other side of the sphere. In the normal direction, the kernel is symmetric with a maximum magnitude slightly larger than the magnitude downstream. The analytical kernel in the Stokes limit possesses spherical symmetry and a maximum magnitude of approximately one third of the actual kernel near the stagnation point. However, the positions of the maxima are approximately similar to those of the actual kernel for $\mathrm{Re}=100$. The maximum of the Gaussian is in the center of the sphere and has a much too large magnitude. It can be concluded that at such small filter widths, the Gaussian is a poor approximation of the regularization of the particle momentum source in the Stokes limit and at larger $\mathrm{Re}$. Note that at finite $\mathrm{Re}$ the particle momentum source does not only have a contribution in the streamwise direction, but also in the normal, or perpendicular, directions. The normal momentum source components have a much smaller magnitude than the streamwise component and their integrals over the entire volume are zero, i.e., there is no resultant momentum source in the normal directions. \\
Figure \ref{fig:momentumsourceshape4Re100} compares the shapes of the regularization of the particle momentum source at $\delta/a=4$. The shape of the kernel is relatively close to a Gaussian that is shifted slightly upstream. Except for this upstream shift, $\mathcal{K}_{\mathrm{St}}$ approximates the shape and the magnitude of the kernel well. Although the Gaussian is much closer to the kernel than for $\delta/a=1$, it still overpredicts the magnitude of the kernel by a factor of approximately $1.5$. \\ 
The studies of the flow relative to a single isolated sphere show that all of the three closure, the viscous closure, the subfilter stress tensor, and the regularization kernel of the particle momentum source, can significantly contribute to the energy transfer and require appropriate modeling in the investigated case. The viscous closure can be obtained analytically and is shown to match the explicit evaluation of $\mathcal{E}_i$. With the non-linear model, the subfilter stresses can be approximated well for small filter widths, but the Smagorinsky model is shown to not be suitable for modeling the subfilter stress tensor. For larger filter widths, $\mathcal{K}_{\mathrm{St}}$ can recover the shape of the regularization of the particle momentum source well. For filter widths $\delta/a>20$, which are relevant for simulations using the PSIC method, conclusions can not be drawn with certainty from the present results. The observed trends suggest, however, that the shape of the regularization kernel approaches a very wide Gaussian and that the subfilter stress tensor still contributes significantly to the energy and momentum transfer, such that it requires modeling. 

\subsubsection{A posteriori analysis}
\begin{figure}
    \centering
    \includegraphics[scale=0.6]{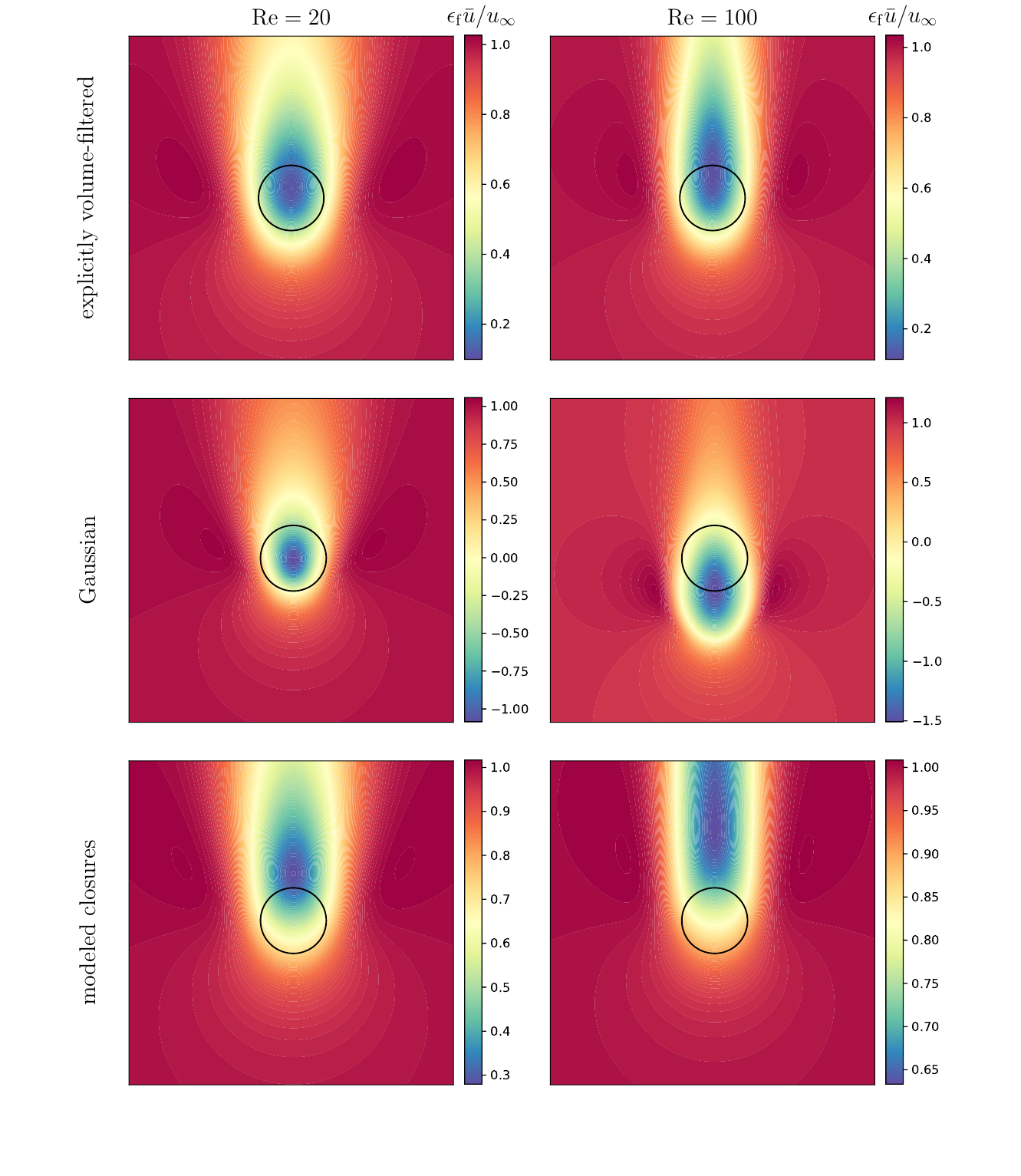}
    \caption{Contours of the filtered streamwise fluid velocity of a flow over a fixed sphere at $\mathrm{Re}=20$ (left) and $\mathrm{Re}=100$ (right). Compared are the explicitly volume-filtered resolved simulation (top), the simplified point-particle framework with a Gaussian regularization and without a model for the subfilter stress tensor (center), and the volume-filtering framework with the analytical kernel of the Stokes regime and the non-linear model for the subfilter stress tensor (bottom). The investigated case corresponds to a filter width $\delta/a=2$.}
    \label{fig:Uaposteriorid2}
\end{figure}
\begin{figure}
    \centering
    \includegraphics[scale=0.6]{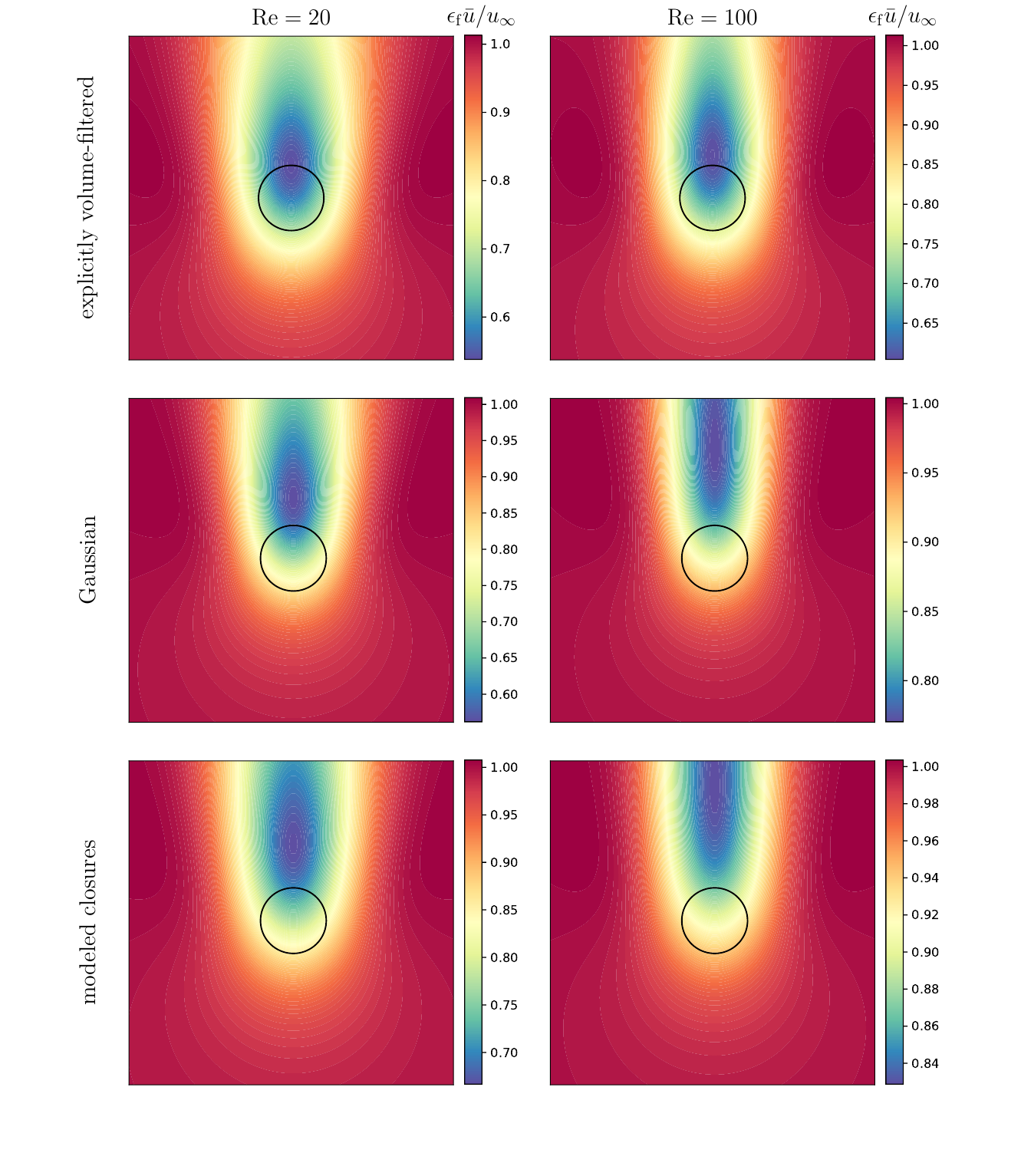}
    \caption{Contours of the filtered streamwise fluid velocity of a flow over a fixed sphere at $\mathrm{Re}=20$ (left) and $\mathrm{Re}=100$ (right). Compared are the explicitly volume-filtered resolved simulation (top), the simplified point-particle framework with a Gaussian regularization and without a model for the subfilter stress tensor (center), and the volume-filtering framework with the analytical kernel of the Stokes regime and the non-linear model for the subfilter stress tensor (bottom). The investigated case corresponds to a filter width $\delta/a=4$.}
    \label{fig:Uaposteriorid4}
\end{figure}

When the volume-filtered equations are solved in a reduced order simulation, such as an Euler-Lagrange or an Euler-Euler simulation, it is particularly relevant what the impact of the closures on the resulting fluid velocity field is. Therefore, we solve the volume-filtered equations with different models for the closures and compare it to the explicitly filtered particle-resolved simulations. To isolate the effect of the closure models on the resulting fluid velocity field, the volume filtered equations are solved on a very fine mesh, i.e., 15 cells per particle diameter, which is much finer than any expected application of volume-filtering. However, this high resolution minimizes the numerical error and the deviations can be traced back to the modeling errors. The simulation setup is similar to the interphase-resolving simulation of the isolated sphere.  \\
In figure \ref{fig:Uaposteriorid2}, we show the contours of the volume-filtered streamwise velocity of the flow past a fixed sphere of $\mathrm{Re}=20$ and $\mathrm{Re}=100$ for three different cases: the explicitly volume-filtered particle-resolved velocity, the Gaussian regularization of the particle momentum source without a model for the subfilter stress tensor, and the regularization with the analytical kernel in the Stokes limit with the non-linear model for the subfilter stress tensor. The results are shown for a filter width of $\delta/a=2$. The case with the Gaussian regularization corresponds to the simplified point-particle framework, which is introduced in section \ref{sec:Stokesflowsinglesphere}. Note that the viscous closure is zero in the present case because the particles are not moving in the considered frame of reference. \\
In figure \ref{fig:Uaposteriorid2}, it can be observed for both $\mathrm{Re}$ that the explicitly volume-filtered fluid velocity possesses a small positive minimal velocity slightly downstream from the particle center. For $\mathrm{Re}=100$ the minimal fluid velocity is further downstream than for $\mathrm{Re}=20$. With the simplified point-particle framework, where a Gaussian is used for regularization of the particle momentum source, a large negative (upstream) velocity is induced for both $\mathrm{Re}$. At the considered filter width, the Gaussian is much too localized or concentrated in the particle center. For $\mathrm{Re}=20$, the minimum streamwise velocity is present at the particle center, and for $\mathrm{Re}=100$ even upstream of the particle. A large spurious velocity is induced that reduces the accuracy of the undisturbed velocity prediction, which is required for the empirical correlations for the drag force and other forces on the particle. A variety of different methods have been developed to subtract this partially spurious self induced fluid velocity \citep{Ireland2017,Balachandar2022,Balachandar2019,Pakseresht2021,Evrard2020a}. With the analytical regularization kernel of the Stokes regime and the non-linear model for the subfilter stress tensor, the minimal velocities are larger positive (downstream) values than for the explicitly volume-filtered velocity and the minima are too far downstream. However, the velocity magnitudes are much closer to the explicitly filtered solution than the velocity of the Gaussian regularization. At $\mathrm{Re}=20$, a good approximation of the explicitly filtered fluid velocity is obtained. For $\mathrm{Re}=100$, we show in section \ref{sec:aprioriparticlemomentumsource} that the shape of the particle momentum source is strongly asymmetric. The assumption of a spherical symmetric kernel of the Stokes regime becomes decreasingly valid for increasing $\mathrm{Re}$. \\
It is worth highlighting that even at the small filter width of $\delta/a=2$ a stable solution of the volume-filtered momentum equation \eqref{eq:momentumwithclosures} together with the volume-filtered continuity equation \eqref{eq:unclosedcontinuity} is obtained. With the single volume fraction advective term, even at larger filter widths numerical interventions are typically required \citep{Link2005,Jing2016}. \\
The results of the same configurations, but with a filter width $\delta/a=4$, are shown in figure \ref{fig:Uaposteriorid4}. At this filter width, the contribution of the subfilter stress tensor is significant and the non-linear model for the subfilter stress tensor predicts too small values. It is shown in section \ref{sec:aprioriparticlemomentumsource}, that  $\mathcal{K}_{\mathrm{St}}$ is a good approximation of the shape of the particle momentum source, even at $\mathrm{Re}=100$, and that the Gaussian is still too localized and possesses a too large magnitude at $\delta/a=4$. \\
The streamwise fluid velocities in figure \ref{fig:Uaposteriorid4} show less deviations than for $\delta/a=2$. The Gaussian regularization captures the magnitude of the minimal velocity for $\mathrm{Re}=20$ well and slightly overestimates it for $\mathrm{Re}=100$. With $\mathcal{K}_{\mathrm{St}}$ and the non-linear model, the minimal streamwise velocity is overestimated more than with the Gaussian regularization for both $\mathrm{Re}$. Although $\mathcal{K}_{\mathrm{St}}$ is a better approximation of the shape of the particle momentum source than the Gaussian, the streamwise fluid velocity field is in better agreement with the explicitly filtered velocity. The reason is that the neglected subfilter stress tensor is at least partially compensated by the overestimation of the regularization kernel. This compensation of errors causes the better agreement when a Gaussian is used at filter width $\delta/a=4$. Instead of relying on compensation of errors, however, the more physical kernel $\mathcal{K}_{\mathrm{St}}$ should be used and a model for the subfilter stress tensor that is also valid at large filter widths needs to be derived. 

\subsection{Periodic array of spheres}
\begin{figure}
    \centering
    \hspace{-1.7cm}
    \includegraphics[scale=0.75]{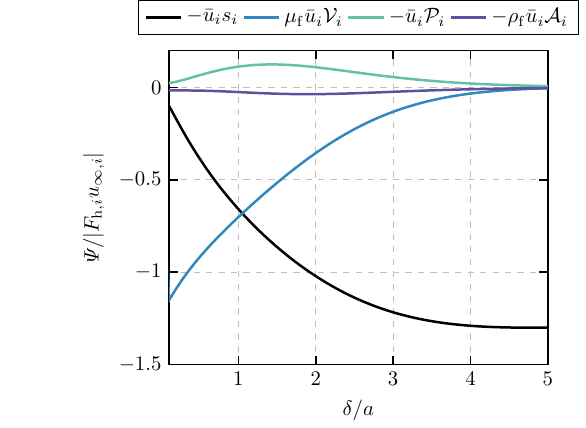}
    \caption{Volume integrated energy transfer rates of the flow over a periodic array of spheres at $\mathrm{Re}_{\mathrm{periodic}}=50$. The energy transfer rates of the particle momentum source, the viscous term, the pressure term, and the advective term are plotted.}
    \label{fig:energytransferperiodicRe50v0}
\end{figure}

\begin{figure}
    \centering
    \hspace{-2.5cm}
    \includegraphics[scale=0.75]{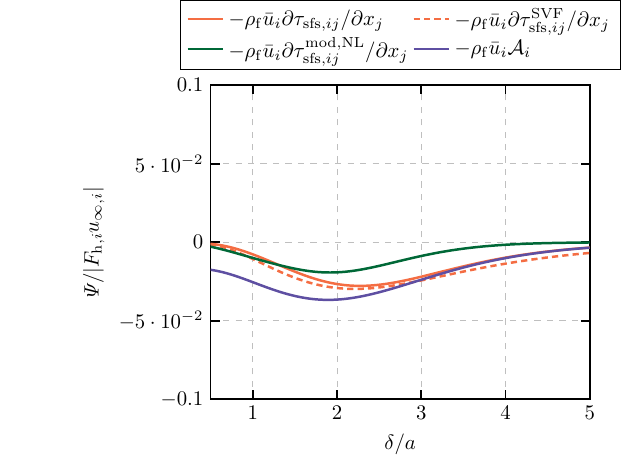}
    \caption{Volume integrated energy transfer rates of the flow over a periodic array of spheres at $\mathrm{Re}_{\mathrm{periodic}}=50$. The energy transfer rates of the advective term, $\rho_{\mathrm{f}}\Bar{u}_i \mathcal{A}_i$, the subfilter stress tensor, $\rho_{\mathrm{f}}\Bar{u}_i \partial \tau_{\mathrm{sfs},ij}/\partial x_j$, the subfilter stress tensor with the single volume fraction definition, $\rho_{\mathrm{f}}\Bar{u}_i \partial \tau_{\mathrm{sfs},ij}^{\mathrm{SVF}}/\partial x_j$, and the modeled subfilter stress tensor with the non-linear model, $\rho_{\mathrm{f}}\Bar{u}_i \partial \tau_{\mathrm{sfs},ij}^{\mathrm{mod,NL}}/\partial x_j$ are plotted.}
    \label{fig:advectiveenergytransferperiodicRe50v0}
\end{figure}

\begin{figure}
    \hspace{0.85cm}
    \includegraphics[scale=0.6]{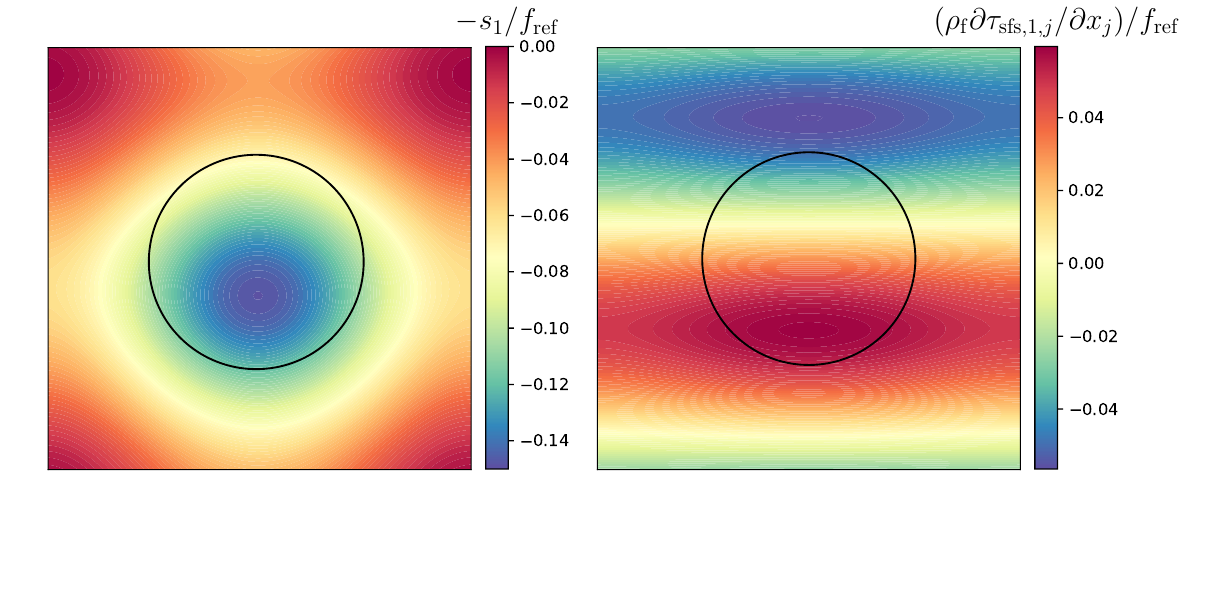}
    \caption{Contours of the momentum contributions in the streamwise direction of the particle momentum source (left) and the divergence of the subfilter stress tensor (right) for a periodic array of spheres at $\mathrm{Re}_{\mathrm{periodic}}=50$ and filter width $\delta/a=4$.}
    \label{fig:momentumtauperiodicRe50v0d4}
\end{figure}

\begin{figure}
    \centering
    \includegraphics[scale=0.75]{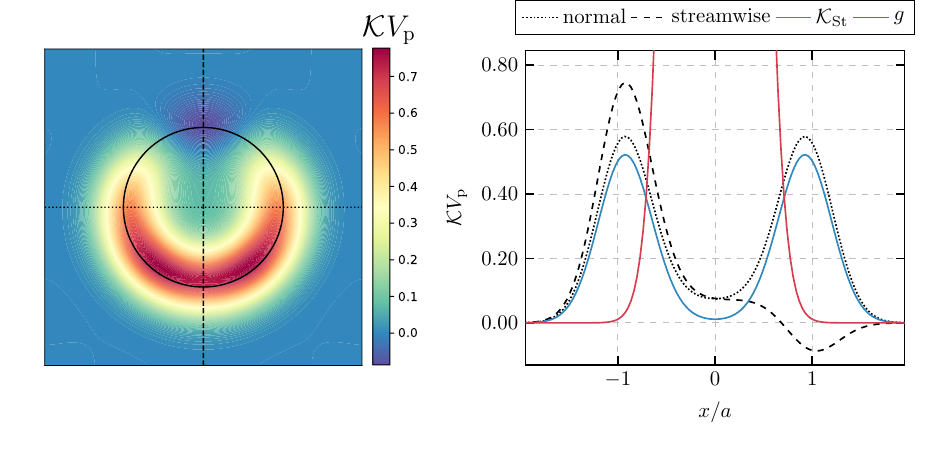}
    \caption{Shape of the particle momentum source for a flow over a fixed array of spheres in the periodic configuration at $\mathrm{Re}_{\mathrm{periodic}}=50$ and filter width $\delta/a=1$. The left plot shows a contour of the regularization kernel. The right plot shows the regularization kernel in the streamwise and normal direction together with the analytical regularization kernel for Stokes flow and a Gaussian of the same filter width.}
    \label{fig:momentumsourceshape1PeriodicRe50}
\end{figure}

\begin{figure}
    \centering
    \includegraphics[scale=0.75]{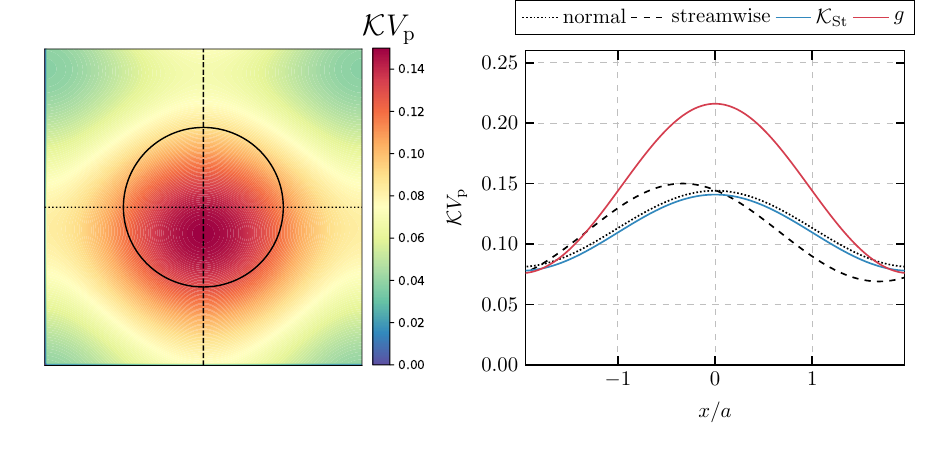}
    \caption{Shape of the particle momentum source for a flow over a fixed array of spheres in the periodic configuration at $\mathrm{Re}_{\mathrm{periodic}}=50$ and filter width $\delta/a=4$. The left plot shows a contour of the regularization kernel. The right plot shows the regularization kernel in the streamwise and normal direction together with the analytical regularization kernel for Stokes flow and a Gaussian of the same filter width.}
    \label{fig:momentumsourceshape4PeriodicRe50}
\end{figure}

To investigate the presence of multiple spheres on the closures, a periodic array of fixed spheres is considered. The flow is driven by a constant momentum source. \\
Figure \ref{fig:energytransferperiodicRe50v0} shows the volume integrated energy transfer rates for the fluid flow through the periodic array of spheres at $\mathrm{Re}_{\mathrm{periodic}}=50$. Since the energy is introduced by the additional driving momentum source, the energy transfer rate of the pressure term is zero at $\delta=0$. The conclusions on the energy transfer are similar to the single sphere configurations, with the exception of the advective term. Because of the periodic domain and the resulting repetition of the flow patterns, all existing flow structures are captured by integrating over a finite volume, whereas in the single sphere configuration a finite integration volume always omits some of the existing flow patterns. The behavior of the advective term with increasing filter width is very similar to the single sphere configuration, but at $\delta=0$ the energy transfer rate of the advective term in the periodic case is zero. In the periodic configuration, the advective term removes energy from large scales and adds it to small scales. However, the contribution of the advective term to the energy transfer is relatively small for all filter widths. \\
In figure \ref{fig:advectiveenergytransferperiodicRe50v0} the energy transfer rates of the subfilter stress tensor, the single volume fraction definition of the subfilter stress tensor, and the modeled subfilter stress tensor using the non-linear model are plotted. The energy transfer rate of the advective term is shown as reference. The energy transfer rate of the present definition and the single volume fraction definition of the subfilter stress tensor are relatively similar and deviate only slightly from each other. The energy transfer rate contribution of the subfilter stress tensor is significant compared to the energy transfer caused by the advective term, especially at large filter widths. Similar to the single sphere cases, the non-linear model for the subfilter stress tensor approximates the energy transfer rate of the subfilter stress tensor well for small filter widths but increasingly deviates from the energy transfer rate of the subfilter stress tensor as the filter width increases. \\
Although the energy transfer rate of the subfilter stress tensor is small compared to the energy transfer rate of the particle momentum source, the subfilter stress tensor can still significantly contribute to the momentum. Figure \ref{fig:momentumtauperiodicRe50v0d4} shows the particle momentum source and the subfilter stress momentum contributions in the streamwise direction for a filter width of $\delta/a=4$. At this filter width, the energy transfer rate of the subfilter stresses is almost zero compared to the energy transfer rate of the particle momentum source. The magnitude of the momentum contribution of the subfilter stresses is approximately one third of the magnitude of the particle momentum source. The significant relative contribution of the subfilter stresses to the fluid momentum confirms that they require modeling even at such large filter widths. \\
As already discussed in the single sphere configuration, another error source in the volume filtered momentum equation is the regularization of the particle momentum source. For the periodic array of spheres at $\mathrm{Re}_{\mathrm{periodic}}=50$, the regularization kernel is shown for $\delta/a=1$ in figure \ref{fig:momentumsourceshape1PeriodicRe50} together with the analytical kernel of the Stokes regime and a Gaussian. Along the streamwise direction the kernel is asymmetric. The largest value of the kernel is at the stagnation point of the sphere. Downstream of the particle the kernel is even negative. In the normal direction the kernel is approximated well by the analytical regularization kernel of the Stokes flow around an isolated sphere, $\mathcal{K}_{\mathrm{St}}$, but the Gaussian of the same filter width possesses a much too large magnitude and wrong shape. \\
The regularization kernel is plotted for $\delta/a=4$ in figure \ref{fig:momentumsourceshape4PeriodicRe50}. The shape of the kernel is relatively similar to a Gaussian in the streamwise and the normal direction but slightly shifted upstream. Due to the periodic domain, the kernel is always significantly larger than zero. At this larger filter width, $\mathcal{K}_{\mathrm{St}}$ is a good approximation of the kernel in terms of shape and magnitude. However, the Gaussian overpredicts the magnitude of the kernel. \\
The main conclusions of the single sphere configuration also apply to the periodic array of spheres. The subfilter stress tensor requires modeling, because it contributes significantly in the momentum equation of the fluid relative to other momentum sources, such as the particle momentum source. For small filter widths, the non-linear model captures the energetic influence of the subfilter stresses. The shape of the regularization of the particle momentum source is poorly approximated by the commonly used Gaussian. At larger filter widths, that are of practical importance for Euler-Lagrange simulations, $\mathcal{K}_{\mathrm{St}}$ approximates the kernel well and is to be preferred over a Gaussian.

\section{Phase-averaged fluid kinetic energy}
\label{sec:fluidKE}
\begin{figure}
    \centering
    \hspace{-0.8cm}
    \includegraphics[scale=0.8]{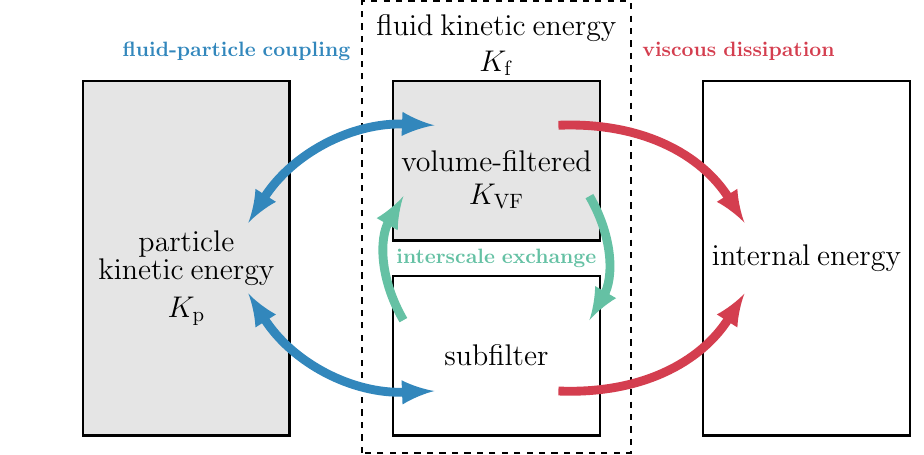}
    \caption{Sketch of the energy transfer in a particle-laden flow without, or only fully elastic, particle collisions. The kinetic energy of the fluid is split into its volume-filtered and subfilter contribution. The arrows indicate energy exchange between fluid and particle, among the fluid scales, and viscous dissipation to internal energy. In the volume-filtered framework, only the boxes highlighted in gray correspond to known quantities. }
    \label{fig:sketchenergytransfer}
\end{figure}

In the literature, numerical frameworks for the transport of gas-solid mixtures or particle-laden flows are commonly assessed by means of the implied fluid-particle energy transfer. We assume a homogeneous suspension and define the phase-averaging operations
\begin{align}
    \langle \varPhi \rangle_{\mathrm{f}} = \dfrac{1}{V_{\mathrm{f}}} \int I_{\mathrm{f}}\varPhi \mathrm{d}V, \\
    \langle \varPhi \rangle_{\mathrm{p}} = \dfrac{1}{V_{\mathrm{p}}} \int (1-I_{\mathrm{f}})\varPhi \mathrm{d}V,
\end{align}
where $\langle \varPhi \rangle_{\mathrm{f}}$ and $\langle \varPhi \rangle_{\mathrm{p}}$ are the fluid and particle phase-averaged quantities and $V_{\mathrm{f}}$ and $V_{\mathrm{p}}$ are the total fluid and particle volume. The total integration volume is given as $V=V_{\mathrm{f}}+V_{\mathrm{p}}$. The transport equations for the phase-averaged fluid kinetic energy $\langle K_{\mathrm{f}} \rangle_{\mathrm{f}}$ and particle kinetic energy $\langle K_{\mathrm{p}} \rangle_{\mathrm{p}}$ are \citep{Subramaniam2014,Mehrabadi2015,Mehrabadi2018}
\begin{align}
\label{eq:fluidkineticenergy}
    \rho_{\mathrm{f}}\dfrac{V_{\mathrm{f}}}{V}\dfrac{\mathrm{d}\langle K_{\mathrm{f}} \rangle_{\mathrm{f}}}{\mathrm{d} t} = -\dfrac{1}{V}\sum_q \int\displaylimits_{\partial\Omega_{\mathrm{p},q}}u_i(-p \delta_{ij} + \mu_{\mathrm{f}} (\dfrac{\partial u_i}{\partial x_j}+\dfrac{\partial u_j}{\partial x_i}))\hat{n}_j\mathrm{d}A - \mathcal{D}, \\
\label{eq:particlekineticenergy}
    \rho_{\mathrm{p}}\dfrac{V_{\mathrm{p}}}{V}\dfrac{\mathrm{d}\langle K_{\mathrm{p}} \rangle_{\mathrm{p}}}{\mathrm{d} t} = \dfrac{1}{V}\sum_q \int\displaylimits_{\partial\Omega_{\mathrm{p},q}}v_{q,i}(-p \delta_{ij} + \mu_{\mathrm{f}} (\dfrac{\partial u_i}{\partial x_j}+\dfrac{\partial u_j}{\partial x_i}))\hat{n}_j\mathrm{d}A,
\end{align}
where $\mathcal{D}$ is the average fluid dissipation and given as
\begin{align}
    \mathcal{D} = \dfrac{1}{V}\int \mu_{\mathrm{f}} I_{\mathrm{f}}\left(\dfrac{\partial u_i}{\partial x_j}\dfrac{\partial u_i}{\partial x_j} + \dfrac{\partial u_i}{\partial x_j}\dfrac{\partial u_j}{\partial x_i} \right)\mathrm{d}V.
\end{align}
The fluid phase-averaging eliminates several terms in the fluid kinetic energy transport equation, the transport or divergence terms, because of the homogeneity of the problem. Because of the no-slip boundary condition equation \eqref{eq:noslip}, the transport equation of the kinetic energy of the mixture $e=\rho_{\mathrm{f}}\dfrac{V_{\mathrm{f}}}{V}K_{\mathrm{f}}+ \rho_{\mathrm{p}}\dfrac{V_{\mathrm{p}}}{V}K_{\mathrm{p}}$ is given as
\begin{align}
\label{eq:mixturekineticenergy}
    \dfrac{\mathrm{d}e}{\mathrm{d}t} = -\mathcal{D}.
\end{align}
Since no inelastic particle collisions are considered, the only mechanism that dissipates energy is the viscous dissipation of the fluid. The validity of equations \eqref{eq:fluidkineticenergy}, \eqref{eq:particlekineticenergy}, and \eqref{eq:mixturekineticenergy} is undisputed \citep{Xu2007a,Subramaniam2014,Mehrabadi2018}. Care has to be taken, however, when the analysis is extended to the volume-filtered framework. \\
An essential modification in the volume-filtered framework is the redefinition of the phase-averaging because, strictly speaking, the fluid and particle phase can not be clearly separated. A phase-averaged volume-filtered quantity is given as
\begin{align}
    \langle \epsilon_{\mathrm{f}} \Bar{\varPhi} \rangle = \dfrac{1}{V_{\mathrm{f}}} \int \epsilon_{\mathrm{f}} \Bar{\varPhi} \mathrm{d}V,
\end{align}
which converges towards $\langle \varPhi \rangle_{\mathrm{f}}$ as $\delta\rightarrow 0$, $\Bar{\varPhi}\rightarrow\varPhi$, and $\epsilon_{\mathrm{f}} \rightarrow I_{\mathrm{f}}$. The phase-averaged volume-filtered kinetic energy transport equation can be obtained by volume-averaging equation \eqref{eq:volumefilteredkineticenergyequation} and possesses several additional non-zero terms 
\begin{align}
\label{eq:phaseaveragedKE}
    \rho_{\mathrm{f}}\dfrac{V_{\mathrm{f}}}{V}\dfrac{\mathrm{d}\langle K_{\mathrm{VF}}\rangle }{\mathrm{d}t} = \hspace{10cm} \nonumber \\
    \dfrac{1}{V}\left(-\rho_{\mathrm{f}}\int\Bar{u}_i \mathcal{A}_i\mathrm{d}V-\int\Bar{u}_i \mathcal{P}_i\mathrm{d}V + \mu_{\mathrm{f}}\int\Bar{u}_i \mathcal{V}_i\mathrm{d}V+\mu_{\mathrm{f}}\int\Bar{u}_i \mathcal{E}_i\mathrm{d}V-\int\Bar{u}_i s_i\mathrm{d}V\right).
\end{align}
In general the right-hand side terms are not zero for $\delta>0$, which is also confirmed by the non-zero energy transfer rates of figure \ref{fig:energytransferperiodicRe50v0}. In the homogeneous case, terms of the form $\int \dfrac{\partial ...}{\partial x_j} \mathrm{d}V$ can be shown to vanish \citep{George2010}. In the volume-filtered framework, terms remain that can not be transformed into such a form. In the literature, the transport equation for the fluid kinetic energy of the phase-averaged volume-filtered or mean fluid velocity is oversimplified by disregarding most of the terms on the right-hand side of equation \eqref{eq:phaseaveragedKE} \citep{Xu2007a,Mehrabadi2015,Subramaniam2014,Mehrabadi2018,Frohlich2018,Keane2023}. However, it can be argued that for spatially constant or small fluid volume fractions, most of the terms on the right-hand side of equation \eqref{eq:phaseaveragedKE} are small, and only the viscous fluid dissipation and the particle momentum source term remain. \\
The energy transfer is visualized in figure \ref{fig:sketchenergytransfer}. In particle-resolved simulations, where the interactions between fluid and particles are fully resolved, the fluid transfers energy to the particles or receives energy from the particles and dissipates energy because of viscosity. In the end, the viscous dissipation is the only mechanism that reduces the total energy $e$ of the fluid-particle mixture. In the volume-filtered framework, however, only the volume-filtered fluid and particle quantities are known, and the subfilter fluid quantities are accounted for by energy exchange terms. The energetic interactions between fluid and particles are only partially resolved as the particles can also exchange energy with the subfilter scales of the fluid. Together with the fact that only a fraction of the viscous dissipation is resolved and that energy can be exchanged with the subfilter scales, this causes the kinetic energy of the particles $K_{\mathrm{p}}$ and the kinetic energy of the volume-filtered fluid $K_{\mathrm{VF}}$ to not balance the resolved viscous dissipation. Therefore, it is not consistent to tailor the closure models in equation \eqref{eq:momentumwithclosures}, especially the shape of the regularization of the particle momentum source, such that $\langle K_{\mathrm{VF}}\rangle = \langle K_{\mathrm{f}} \rangle_{\mathrm{f}}$ \citep{Xu2007a,Evrard2020a}. Specific regularization kernels of the particle momentum source have been derived that produce an artificial and purely numerical dissipation, such that underresolved viscous dissipation is compensated \citep{Evrard2020a,Keane2023}. Instead of empirically varying the regularization kernel to achieve $\langle K_{\mathrm{VF}}\rangle = \langle K_{\mathrm{f}} \rangle_{\mathrm{f}}$, we suggest to explicitly volume-filter the existing particle-resolved DNS of, e.g., \citet{Mehrabadi2018} or \citet{Frohlich2018} with different filter widths as an extension of the present study. \\


\section{Guidelines for point-particle simulations}
\label{sec:guidelines}

\subsection{Choosing an appropriate filter width}
The volume-filtered NSE, including the derived closures, can be solved in the scope of a point-particle simulation, where the influence of the particles on the fluid is considered by the regularized particle-momentum source. Point-particle simulations are typically carried out on a computational fluid mesh that is coarse compared to the size of the particles and, in addition to the modeling errors discussed in the previous sections, discretization errors occur. \\
There are three length scales in point-particle simulation that have to be considered, the particle diameter, $d_{\mathrm{p}}$, the size of the computational fluid mesh cells, $\Delta x$, and the filter width, $\delta$. Typically, the particle diameter is predetermined and the remaining parameters have to be chosen accordingly. It is demonstrated in the previous sections, that a Gaussian kernel approximates the particle momentum source well, even for Reynolds numbers up to $\mathrm{Re}=100$, if the filter width is large enough. In Stokes flow, a filter width of $\delta/d_{\mathrm{p}} \gtrapprox 3.5$ yields an analytical kernel that deviates insignificantly from a Gaussian. A Gaussian kernel is advantageous, as it can be easily integrated over a cell of the fluid mesh. As shown by \citet{Balachandar2022}, at large filter widths the following approximation becomes valid 
\begin{align}
    \epsilon_{\mathrm{p}}(\boldsymbol{x}) = \int\displaylimits_{\Omega} I_{\mathrm{p}}(\boldsymbol{y})g(|\boldsymbol{x}-\boldsymbol{y}|)\mathrm{d}V_y \approx V_{\mathrm{p}}g(|\boldsymbol{x}|), \quad \sigma\gg d_{\mathrm{p}}
\end{align}
where $I_{\mathrm{p}}$ is the particle indicator function and $\sigma$ is the standard deviation of the Gaussian. For $\sigma/d_{\mathrm{p}}=1$, which corresponds to $\delta/d_{\mathrm{p}}\approx3.76$, this approximation leads to a maximum deviation of $7.7\%$ from the analytical volume fraction. To justify the Gaussian approximation of the particle momentum source and the volume fraction, we suggest to choose the filter width at least $\delta/d_{\mathrm{p}}>3.76$ or $\sigma/d_{\mathrm{p}}>1$ independent of the fluid mesh size. \\
The filter width can not be chosen independently of the mesh size. The purpose of solving the volume-filtered NSE instead of the NSE is that large gradients that can not be resolved by a coarse fluid mesh are avoided. In the volume-filtered NSE, large gradients are mainly avoided by choosing a widely distributed particle momentum source. If the size of the fluid mesh cell is too coarse for the chosen filter width, the discretization error imposes a spatially varying unknown filter, and the closures, which are derived for a spatially constant Gaussian filter of a specific filter width, are not valid. If the filter width is chosen $\sigma/\Delta x \gtrapprox 1$, a Gaussian can be well resolved by the mesh. Sampling a Gaussian on a mesh with a significantly smaller ratio of $\sigma/\Delta x$ may capture only a fraction of its energy. \\
For point-particle simulations, we recommend to choose the filter width based on what is larger, the particle diameter, $d_{\mathrm{p}}$, or the fluid mesh size, $\Delta x$. The filter width should satisfy both, $\sigma/d_{\mathrm{p}} \gtrapprox 1$ and $\sigma/\Delta x \gtrapprox 1$, to justify the Gaussian approximation of the volume fraction and the particle momentum source and to avoid the solution being altered by the filter imposed by a too coarse fluid mesh. \\
Note that the guidelines also hold in a turbulent flow independent of the size of the Kolmogorov length scale. In an LES, even without particles, $\sigma/\Delta x \gtrapprox 1$ should be satisfied to avoid flow structures that are too small to be resolved by the mesh.

\subsection{Implementation of the closures}
\begin{figure}
    \centering
    \includegraphics[scale=0.6]{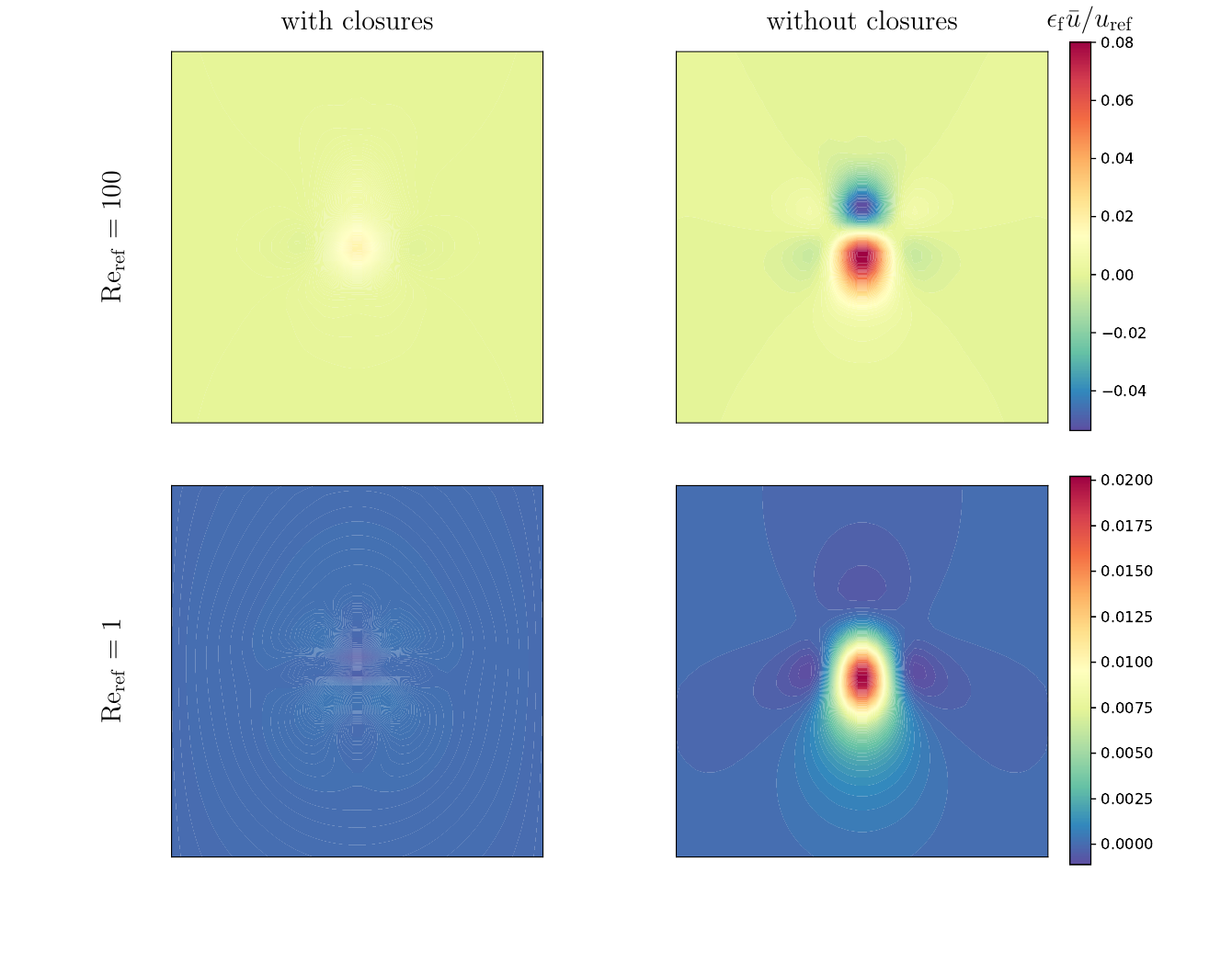}
    \caption{Contours of the spurious streamwise fluid velocity in a configuration without relative velocity between fluid and particle in a frame of velocity $u_{\mathrm{ref}}$. The spurious currents resulting from the volume-filtered NSE with and without the closures $\tau_{\mathrm{sfs},ij}^{\mathrm{G}}$ and $\mathcal{E}_i$ are compared for $\mathrm{Re}_{\mathrm{ref}}=100$ and $\mathrm{Re}_{\mathrm{ref}}=1$ with a resolution $d_{\mathrm{p}}/\Delta x = 2$ and a filter width $\sigma/d_{\mathrm{p}}=1$. }
    \label{fig:GalileanInvariancefine}
\end{figure}

\begin{figure}
    \centering
    \includegraphics[scale=0.6]{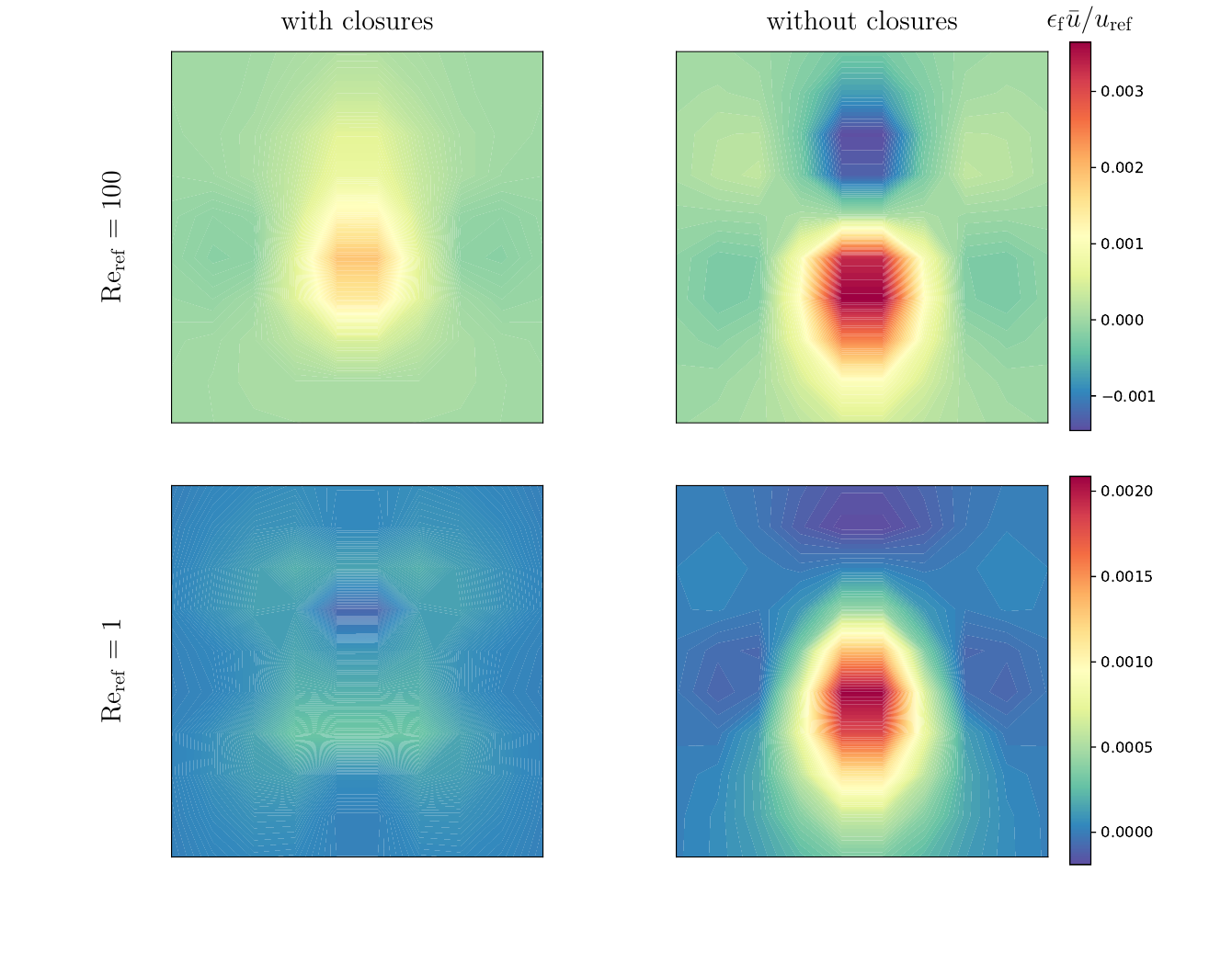}
    \caption{Contours of the spurious streamwise fluid velocity in a configuration without relative velocity between fluid and particle in a frame of velocity $u_{\mathrm{ref}}$. The spurious currents resulting from the volume-filtered NSE with and without the closures $\tau_{\mathrm{sfs},ij}^{\mathrm{G}}$ and $\mathcal{E}_i$ are compared for $\mathrm{Re}_{\mathrm{ref}}=100$ and $\mathrm{Re}_{\mathrm{ref}}=1$ with a resolution $d_{\mathrm{p}}/\Delta x = 0.5$ and a filter width $\sigma/d_{\mathrm{p}}=2$. }
    \label{fig:GalileanInvariancecoarse}
\end{figure}

In the volume-filtered continuity equation \eqref{eq:divergencevolumefiltered}, in the Galilean invariance part of the subfilter stress tensor $\tau_{\mathrm{sfs},ij}^{\mathrm{G}}$, and in the viscous closure $\mathcal{E}_i$, spatial gradients of the volume-fraction are required. In a finite volume framework, the spatial gradients are required as integrals over a fluid mesh cell. If the volume fraction can be approximated with a Gaussian, the following integrals over a fluid mesh cell $\Omega_{\Delta}$ can be computed analytically as
\begin{align}
    \int_{\Omega_{\Delta}}\dfrac{\partial \epsilon_{\mathrm{p}}}{\partial x_i}\mathrm{d}V &\approx V_{\mathrm{p}} \int_{\Omega_{\Delta}}\dfrac{\partial g}{\partial x_i}\mathrm{d}V, \\
    \int_{\Omega_{\Delta}}\dfrac{\partial^2 \epsilon_{\mathrm{p}}}{\partial x_i\partial x_i}\mathrm{d}V &\approx V_{\mathrm{p}} \int_{\Omega_{\Delta}}\dfrac{\partial^2 g}{\partial x_i\partial x_i}\mathrm{d}V, \\
    \int_{\Omega_{\Delta}}\dfrac{\partial \epsilon_{\mathrm{p}}^2}{\partial x_i}\mathrm{d}V &\approx V_{\mathrm{p}}^2 \int_{\Omega_{\Delta}}\dfrac{\partial g^2}{\partial x_i}\mathrm{d}V.
\end{align}
An accurate estimate of the spatial gradients of the volume-fraction in the continuity equation, in the closure $\tau_{\mathrm{sfs},ij}^{\mathrm{G}}$, and in the viscous closure $\mathcal{E}_i$ is crucial to achieve Galilean invariance in the numerical simulations. Galilean invariance is satisfied if the following relation for the volume-filtered velocity in the new frame of reference $\epsilon_{\mathrm{f}}\Bar{\mathcal{U}}_i$ holds
\begin{align}
\label{eq:volumefilteredGalileaninvariance}
    \epsilon_{\mathrm{f}}\Bar{\mathcal{U}}_i(\boldsymbol{x},t) &= \int\displaylimits_{\Omega} I_{\mathrm{f}}(\boldsymbol{y}) (u_i(\boldsymbol{y}-\boldsymbol{u}_{\mathrm{ref},i}t,t) + u_{\mathrm{ref},i})g(|\boldsymbol{x}-\boldsymbol{y}|)\mathrm{d}V_y \nonumber \\
    &= \epsilon_{\mathrm{f}}\Bar{u}_i(\boldsymbol{x}-\boldsymbol{u}_{\mathrm{ref},i}t,t) + \epsilon_{\mathrm{f}}u_{\mathrm{ref},i}(\boldsymbol{x},t).
\end{align}
To isolate the effect of determining the terms involving spatial gradients of the volume-fraction, a fixed particle in a quiescent fluid is considered. The particle-momentum source is zero in this case. Transforming the frame of reference with a velocity $u_{\mathrm{ref}}$, yields a fluid velocity and a particle velocity equal to $u_{\mathrm{ref}}$ in the new frame. The volume-filtered fluid velocity is given according to equation \eqref{eq:volumefilteredGalileaninvariance}. Any deviation from $\epsilon_{\mathrm{f}}\Bar{\mathcal{U}}_i$ of the volume-filtered velocity in the new frame is considered a spurious current. Without the terms $\tau_{\mathrm{sfs},ij}^{\mathrm{G}}$ and $\mathcal{E}_i$ in the volume-filtered momentum equation, spurious currents occur because 
$ \epsilon_{\mathrm{f}}\Bar{u}_i$ is not divergence-free. Based on the velocity of the moving frame of reference, we consider two Reynolds numbers, $\mathrm{Re}_{\mathrm{ref}}=u_{\mathrm{ref}}d_{\mathrm{p}}\rho_{\mathrm{f}}/\mu_{\mathrm{f}}$. For $\mathrm{Re}_{\mathrm{ref}}=100$ the closure $\tau_{\mathrm{sfs},ij}^{\mathrm{G}}$ is dominant and for $\mathrm{Re}_{\mathrm{ref}}=1$ the viscous closure $\mathcal{E}_i$ dominates. Figure \ref{fig:GalileanInvariancefine} shows the spurious currents when the frame of reference is transformed using the proposed closures computed as described and without the closures and a resolution of $d_{\mathrm{p}}/\Delta x = 2$. The filter with is chosen according to the guidelines $\sigma/d_{\mathrm{p}}=1$. Without the closures $\tau_{\mathrm{sfs},ij}^{\mathrm{G}}$ and $\mathcal{E}_i$, a spurious velocity of up to $8\%$ of $u_{\mathrm{ref}}$ occurs. With the proposed discretization of the volume fraction gradients, the closures drastically reduce the spurious currents at this resolution. Note that the closures are exact and solely the discretization causes the observed small spurious currents. Figure \ref{fig:GalileanInvariancecoarse} shows the spurious currents at a coarser resolution of $d_{\mathrm{p}}/\Delta x = 0.5$ and a filter width $\sigma/\Delta x = 1$, according to the guidelines. The maximum magnitude of the spurious velocity is relatively small even without closures. The closures, however, reduce the spurious velocity even further. The spurious currents reduce when the filter width is increased but they can reach a significant magnitude even if the filter with is large. In configurations with a large mean flow velocity and relatively small relative velocity between fluid and particles, such as channel flows, $u_{\mathrm{ref}}$ is large and can induce spurious currents that are large compared to the velocity induced by the particle momentum source. \\
The present test case demonstrates that an additional viscosity, as applied by \citet{Zhang1997a,Patankar2001a,Capecelatro2013}, is not suitable to replace the viscous closure, as it solely decreases $\mathrm{Re}_{\mathrm{ref}}$, which does not avoid the spurious currents. \\
It is shown in the previous sections, that the subfilter stress tensor can significantly contribute to the energy transfer and the fluid momentum balance for small and large filter widths and, therefore, requires appropriate modeling. The proposed non-linear model for the subfilter stress tensor predicts excellent agreement with the subfilter stress tensor for small filter widths, but for large filter widths large deviations are observed. A Smagorinsky type model for the subfilter stress tensor can not predict the energy transfer of the subfilter stress tensor, independent of the model constant. For filter widths according to the guidelines, neither of the discussed models for $\tau_{\mathrm{sfs},ij}$ can be recommended for point-particle simulation and further research is required to derive models that are valid for large filter widths. \\
The closures that are discussed in the present paper including their expressions, their implementation and the conditions under which they are valid are summarized in table \ref{tab:summary}. 
\begin{table}
  \begin{center}
\def~{\hphantom{0}}
  \begin{tabular}{p{1.5cm}p{2.2cm}p{4.5cm}p{4.5cm}}
      Closure & Expression & Implementation & Validity \\[3pt]
       $\mathcal{E}_i$   & equation \eqref{eq:viscousclosure} & Gaussian approximation of $\epsilon_{\mathrm{p}}$ and analytical integration over fluid mesh cells & spherical or non-rotating non-spherical particles\\
       $s_{q,i}$ & $F_{\mathrm{h},q,i}g$ & $F_{\mathrm{h},q,i}$ from empirical correlation, analytical integration of Gaussian over fluid mesh cells & $\delta/d_{\mathrm{p}} \gtrapprox 3.5$\\
       $\tau_{\mathrm{sfs},ij}$ & equation \eqref{eq:nonlinearmodel} & Gaussian approximation of $\epsilon_{\mathrm{p}}$ and analytical integration over fluid mesh cells & Galilean invariance contribution is exact, non-linear model accurate up to $\delta/d_{\mathrm{p}}\approx 1$
  \end{tabular}
  \caption{Summary of the closures discussed in the present paper, how they can be implemented in point-particle simulations, and under which conditions they are valid.}
  \label{tab:summary}
  \end{center}
\end{table}

\subsection{Remark on the PSIC method}
With the very popular PSIC method \citep{Crowe1977}, the particle momentum source is applied to the computational cell in which the particle lies, which may satisfy $\sigma/d_{\mathrm{p}} \gtrapprox 1$ for small particles, but always violates $\sigma/\Delta x \gtrapprox 1$. Therefore, the PSIC method is not in accordance with the guidelines for the filter width. This can have at least the following potentially problematic consequences: (\romannumeral 1) If particles move from one fluid mesh cell to another, sudden jumps in the particle momentum source may occur. (\romannumeral 2) A fluid mesh cell may contain significantly more particles than a neighboring mesh cell, which can cause large gradients of the particle momentum source and, consequently, large fluid velocity gradients that can not be resolved by the fluid mesh. The filter can not be assumed to be a spatially uniform Gaussian and additional unknown closures arise. (\romannumeral 3) The PSIC method proposed by \citet{Crowe1977} leads to a regularization of the particle momentum source and, consequently, an induced local fluid velocity that depends on the ratio $d_{\mathrm{p}}/\Delta x$. In point-particle simulations, the local fluid velocity interpolated to the particle position is used to compute the drag force on the particle. As shown by \citet{Evrard2021}, the PSIC method can cause significant local fluid velocity disturbances even for $d_{\mathrm{p}}/\Delta x\approx 0.1$, which leads to an erroneous prediction of the forces acting on the particle. (\romannumeral 4) Since the local filter width implied by the PSIC method depends on the local mesh resolution, local mesh refinement, e.g., near a solid wall, leads to a non-uniform filter and additional closures that are neglected. \\
\citet{Eaton2009} summarizes commonly observed discrepancies between simulations using the PSIC method and experiments, which are potentially caused by these problems. Regularizing the particle momentum source over multiple cells using a Gaussian with a sufficiently large spatially uniform standard deviation can mitigate the mentioned problems of the PSIC method. However, the complexity of the implementation is increased and some computational overhead is added. 


\section{Conclusions}
\label{sec:conclusions}
In the present paper, we investigate the closures that arise in the fluid momentum equation when the Navier-Stokes equations describing an incompressible particle-laden flow are volume-filtered. Closures are required for the viscous closure arising from volume-filtering the viscous term in the NSE, the regularization of the particle momentum source and the subfilter stress tensor that arises from the volume-filtering the advective term in the NSE. A new form of the advective term in the volume-filtered momentum equation is proposed that circumvents frequently reported stability issues for locally small fluid volume fractions. \\
In this paper, an analytical expression is derived for the viscous closure that is valid without introducing further assumptions. Therefore, this term can be considered closed. For the case of an isolated sphere in the Stokes regime, we derive an analytical expression for the particle momentum source. We show that the commonly used Gaussian regularization fundamentally deviates from the analytical expression in shape and magnitude for small filter widths, but converges to the analytical expression of the particle momentum source in Stokes regime with increasing filter width. Particle-resolved simulations of isolated spheres at finite $\mathrm{Re}$ and the flow through a periodic array of spheres reveal that the analytical regularization kernel in the Stokes limit requires adjustment for small filter widths, but is a good approximation for larger filter widths. With the same particle-resolved simulations, we show that the subfilter stress tensor contributes significantly to the energy transfer and momentum balance and, contrary to the common practice, requires modeling. We derive the non-linear model for volume-filtered particle-laden flows and adapt the Smagorinsky model to volume-filtered particle-laden flows. The non-linear model predicts the subfilter stress tensor well for small filter widths, but becomes an increasingly worse approximation for larger filter widths, whereas the adapted Smagorinsky model can not reproduce the right energy transfer. \\
Simulations of the isolated sphere configuration at finite $\mathrm{Re}$ are carried out, where instead of resolving the particle surface, the volume-filtered Navier-Stokes equations with the modeled closures are solved. We observe that a Gaussian regularization of the particle momentum source induces a large spurious velocity at small filter widths. Although the analytical regularization kernel in the Stokes limit is a good approximation of the particle momentum source at larger filter widths, deviations of the velocity field occur because of the lack of a reliable model for the subfilter stress tensor at large filter widths. \\
With the theoretical and empirical findings of the present paper, we show that the common equation for the phase-averaged kinetic energy of the volume-filtered velocity is not accurate and includes additional terms that contribute to the energy transfer between filtered scales and subfilter scales. \\
Finally, guidelines for how to choose the filter width in point-particle simulations are deduced from the new findings. We suggest an accurate implementation of the proposed closures at resolutions typical for point-particle simulations and demonstrate that if the closures are neglected, Galilean invariance is violated. We identify problems that can occur when applying the very popular PSIC method that may cause the observed discrepancies between simulations and experiments. Following the proposed guidelines together potentially mitigates the identified problems of simulations using the PSIC method.






\appendix

\section{Contribution of pure particle rotation to the viscous closure}
\label{ap:rotationviscouclosure}
We decompose the fluid velocity at the particle surface into a pure translation with the velocity $\boldsymbol{v}_{q}$ and a pure rotation of the particle with the angular velocity $\boldsymbol{\omega}_q$
\begin{align}
    \boldsymbol{u}|_{\partial \Omega_{\mathrm{p},q}} = \boldsymbol{v}_{q} + \boldsymbol{\omega}_q \times (\boldsymbol{y}-\boldsymbol{x}_{\mathrm{p},q}),
\end{align}
where $\boldsymbol{x}_{\mathrm{p},q}$ is the center of the particle. Convolution over the particle surface according to equation \eqref{eq:switichingfilteringderivative} gives
\begin{align}
    \sum\displaylimits_q\int\displaylimits_{\partial \Omega_{\mathrm{p},q}}g(|\boldsymbol{x-y}|) (\boldsymbol{\omega}_q\times(\boldsymbol{y}-\boldsymbol{x}_{\mathrm{p},q})) \cdot \hat{\boldsymbol{n}} dA_y = 0,
\end{align}
because $(\boldsymbol{y}-\boldsymbol{x}_{\mathrm{p},q}) || \hat{\boldsymbol{n}}$. Therefore, the rotation of a spherical particle does not contribute to the viscous closure.

\section{Subfilter stress tensor equation}
\label{ap:subfilterstressequation}
Consider the derivative of the squared volume-filtered fluid velocity with respect to $\sigma^2$. By inserting equation \eqref{eq:differentialfiltering}, we obtain
\begin{align}
\label{eq:duudsigma}
    \dfrac{\partial \overline{I_{\mathrm{f}}u_i}\vert_{\sigma}\overline{I_{\mathrm{f}}u_j}\vert_{\sigma}}{\partial (\sigma^2)} = \overline{I_{\mathrm{f}}u_j}\vert_{\sigma} \dfrac{\partial \overline{I_{\mathrm{f}}u_i}\vert_{\sigma}}{\partial (\sigma^2)} + \overline{I_{\mathrm{f}}u_i}\vert_{\sigma} \dfrac{\partial \overline{I_{\mathrm{f}}u_j}\vert_{\sigma}}{\partial (\sigma^2)} \nonumber \\ = \dfrac{1}{2}\left[ \overline{I_{\mathrm{f}}u_j}\vert_{\sigma} \dfrac{\partial^2 \overline{I_{\mathrm{f}}u_i}\vert_{\sigma}}{\partial x_k \partial x_k} + \overline{I_{\mathrm{f}}u_i}\vert_{\sigma} \dfrac{\partial^2 \overline{I_{\mathrm{f}}u_j}\vert_{\sigma}}{\partial x_k \partial x_k}\right].
\end{align}
Furthermore, the Laplacian of the squared volume-filtered fluid velocity can be expanded
\begin{align}
\label{eq:d2udxdx}
    \dfrac{\partial}{\partial x_k}\left( \dfrac{\partial \overline{I_{\mathrm{f}}u_i}\vert_{\sigma} \overline{I_{\mathrm{f}}u_j}\vert_{\sigma} }{\partial x_k} \right) = \dfrac{\partial}{\partial x_k}\left(\overline{I_{\mathrm{f}}u_j}\vert_{\sigma} \dfrac{\partial \overline{I_{\mathrm{f}}u_i}\vert_{\sigma}}{\partial x_k} + \overline{I_{\mathrm{f}}u_i}\vert_{\sigma} \dfrac{\partial \overline{I_{\mathrm{f}}u_j}\vert_{\sigma}}{\partial x_k} \right) \\ \nonumber
    = \overline{I_{\mathrm{f}}u_j}\vert_{\sigma} \dfrac{\partial ^2\overline{I_{\mathrm{f}}u_i}\vert_{\sigma}}{\partial x_k \partial x_k} + \overline{I_{\mathrm{f}}u_i}\vert_{\sigma} \dfrac{\partial^2 \overline{I_{\mathrm{f}}u_j}\vert_{\sigma}}{\partial x_k \partial x_k} + 2 \dfrac{\partial \overline{I_{\mathrm{f}}u_i}\vert_{\sigma}}{\partial x_k}\dfrac{\partial \overline{I_{\mathrm{f}}u_j}\vert_{\sigma}}{\partial x_k}.
\end{align}
Equation \eqref{eq:duudsigma} and equation \eqref{eq:d2udxdx} can be combined to obtain
\begin{align}
    \dfrac{\partial \overline{I_{\mathrm{f}}u_i}\vert_{\sigma} \overline{I_{\mathrm{f}}u_j}\vert_{\sigma}}{\partial (\sigma^2)} = \dfrac{1}{2}\left[ \dfrac{\partial^2 \overline{I_{\mathrm{f}}u_i}\vert_{\sigma}\overline{I_{\mathrm{f}}u_j}\vert_{\sigma}}{\partial x_k \partial x_k} - 2 \dfrac{\partial \overline{I_{\mathrm{f}}u_i}\vert_{\sigma}}{\partial x_k}\dfrac{\partial \overline{I_{\mathrm{f}}u_j}\vert_{\sigma}}{\partial x_k}\right].
\end{align}
Together with the definition of the subfilter stress tensor 
\begin{align}
    \tau_{\mathrm{sfs},ij}\vert_{\sigma} = \overline{I_{\mathrm{f}}u_iu_j}\vert_{\sigma} - \overline{I_{\mathrm{f}}u_i}\vert_{\sigma}\overline{I_{\mathrm{f}}u_j}\vert_{\sigma},
\end{align}
and equation \eqref{eq:differentialfiltering} applied to the volume-filtered squared fluid velocity
\begin{align}
    \dfrac{\partial \overline{I_{\mathrm{f}}u_iu_j}\vert_{\sigma}}{\partial (\sigma^2 )} = \dfrac{1}{2}\nabla^2\overline{I_{\mathrm{f}}u_iu_j}\vert_{\sigma},
\end{align}
the equation for the subfilter stress tensor is obtained
\begin{align}
    \dfrac{\partial \tau_{\mathrm{sfs},ij}\vert_{\sigma}}{\partial (\sigma^2)} = \dfrac{1}{2}\nabla^2\tau_{\mathrm{sfs},ij}\vert_{\sigma} + \dfrac{\partial \overline{I_{\mathrm{f}}u_i}\vert_{\sigma}}{\partial x_k}\dfrac{\partial \overline{I_{\mathrm{f}}u_j}\vert_{\sigma}}{\partial x_k}.
\end{align}

\bibliographystyle{jfmc}

\end{document}